\documentclass[
 reprint,
superscriptaddress,
 amsmath,amssymb,
 aps,
prb,
floatfix,
]{revtex4-2}

\usepackage{natbib}
\setcitestyle{square, comma, numbers,sort&compress, super}
\usepackage{bbold}
\usepackage{esint}
\usepackage{amssymb}
\usepackage{mathtools}
\usepackage{xcolor}
\usepackage{physics}
\usepackage{float}
\usepackage[caption=false]{subfig}
\usepackage{graphicx}
\usepackage{dcolumn}
\usepackage{bm}
\usepackage{hyperref}

\usepackage[normalem]{ulem}

\begin{document}

\preprint{APS/123-QED}

\title{Thermalization of Non-Fermi Liquid Electron-Phonon Systems:\\ Hydrodynamic Relaxation of the Yukawa-SYK Model}

\author{Hossein Hosseinabadi}
\email[hhossein@uni-mainz.de]{}
\affiliation{Institut für Physik, Johannes Gutenberg-Universität Mainz, 55099 Mainz, Germany}
\author{Shane P. Kelly}
\affiliation{Institut für Physik, Johannes Gutenberg-Universität Mainz, 55099 Mainz, Germany}
\author{Jörg Schmalian}
\affiliation{Institut für Theorie der Kondensierten Materie, Karlsruher Institut für Technologie, 76131 Karlsruhe, Germany}
\affiliation{Institut für Quantenmaterialien und Technologien, Karlsruher Institut für Technologie, 76131 Karlsruhe, Germany}
\author{Jamir Marino}
\affiliation{Institut für Physik, Johannes Gutenberg-Universität Mainz, 55099 Mainz, Germany}
\date{\today}

\begin{abstract}
We study thermalization dynamics in a fermion-phonon variant of the Sachdev-Ye-Kitaev model coupled to an external cold thermal bath of harmonic oscillators. 
We find that quantum critical fermions thermalize more efficiently than phonons, in sharp contrast to the behavior in the Fermi liquid regime. 
In addition, after a short prethermal stage, the system acquires a quasi-thermal distribution given by a time-dependent effective temperature, reminiscent of "hydrodynamic" relaxation. All physical observables relax at the same rate which scales with the final temperature through an exponent that  depends universally on the low energy spectrum of the system and the bath. Such relaxation rate is derived using a hydrodynamic approximation in full agreement with the numerical solution of a set quantum kinetic equations derived from the Keldysh formalism for non-equilibrium Green's functions. Our results hint toward further research on the applicability of the hydrodynamic picture in the description of the late time dynamics of open quantum systems despite the absence of conserved quantities in regimes dominated by conserving collisions.
\end{abstract}

\maketitle

\section{Introduction}
Thermalization of isolated and open quantum many-body systems has been a subject of intense research in the last decade \cite{Abanin_RMP19,Mori_2018,Marino_2022,Vega_RMP17}. The combination of the large number of degrees of freedom and extended correlations in these systems restricts the applicability of numerical and perturbative methods. Not suffering from these limitations, exactly solvable models can be valuable platforms for the study of relaxation in quantum many-body systems. Instances of these systems include integrable models \cite{Rigol_PRL07,Pozsgay_PRL14,Wouters_PRL14,Vidmar_2016} and the large $N$ limit of quantum $O(N)$ field theories \cite{Berges_NuclPA02,Berges_PRL08,Chiocchetta_PRL17} featuring ergodicity breaking and the phenomenon of prethermalization \cite{Berges_PRL04,Langen_2016}.

A pivotal case of exact solvability is the celebrated Sachdev-Ye-Kitaev model (SYK) \cite{sachdev_ye,kitaev,complex_syk,sachdev_rmp} consisting of a system of randomly interacting Majorana fermions which admits an exact solution in the limit of a very large number of fermions. The SYK model features a low temperature non Fermi liquid (NFL) critical state, Planckian dissipation together with the saturation of the upper bound on quantum chaos \cite{Maldacena-prd16,stanford-shenker}. Last but not least, it is shown that the SYK model has a holographic dual as a theory of gravity in AdS background \cite{kitaev,Sachdev_PRX15,Jevicki_JHEP16}. Another class of SYK models termed as Yukawa-SYK models, comprises systems of complex fermions interacting randomly with phonons \cite{ysyk2019,wang-prl2020,Wang_Chubokov_2020,classen-chubukov,davis2022} or similar bosonic excitations \cite{Wang-prb21} with a plethora of interesting properties in addition to those of purely fermionic SYK systems. These include a NFL to unconventional superconducting phase transition \cite{ysyk2019,HAUCK2020,wang-prl2020,classen-chubukov,Inkof_Nat22} beyond the BCS theory \cite{BCS} and self-tuned quantum criticality for phonons where the phonon gap vanishes at $T\to 0$ regardless \cite{ysyk2019,wang-prl2020,Wang_Chubokov_2020,classen-chubukov,Pan-prr21} of the strength of the interaction and the phonon gap. Extensions of these models to lattice systems \cite{Chowdhurry_PRX18,Kim_2021,Esterlis_1,Esterlis_2,Tikhanovskaya_PRL22} and to the Kondo problem \cite{Jang_2023} have also been studied in the past.

In this paper, we study the dynamics of a Yukawa-SYK system coupled to an external bath of thermal phonons by using the Keldysh formalism of non-equilibrium quantum field theory \cite{Keldysh,Kadanoff,kamenev}. Previous works have addressed various out of equilibrium aspects of isolated \cite{eberlein,Bhattarcharya,Haldar_PRR2020,schiro-prb22}, open \cite{almheir-syk-2019,zhang-syk-2019,cheipesh2021} and periodically driven \cite{knapp-prl20} fermionic SYK models. Our motivation for considering the dynamics of Yukawa-SYK model is threefold: (i) Phonons and other bosonic excitations are always present in real electronic systems and therefore, the Yukawa-SYK model is a convenient framework to study their effects on the NFL behavior. As we will see, the model is an example for the complex non-equilibrium dynamics of a multi-component systems with strongly-interacting parts. (ii) Most of the studies of the dynamics of open SYK systems consider critical NFL fermionic baths which usually are modeled by another SYK system ,while in realistic  settings, there are no such restrictions on the nature of the external bath. Instead, in this work we consider a generic Caldeira-Leggett \cite{caldeira} bath of harmonic oscillators coupled to the phonons of the system. Below we demonstrate that this is indeed the most relevant bath coupling of our system. (iii) The fermion sector of the Yukawa-SYK model has a $U(1)$ symmetry and has a finite local Hilbert space while these features are absent in the phonon sector. The $U(1)$ symmetry of the Yukawa-SYK model can break, yielding a superconducting state. Thus, our study of the symmetric limit is a necessary first step to non-equilibrium superconductivity \cite{babadi_prb16,babadi_prb17,Foster_PRB17,Murakami_PRB17,OjedaCollado_PRB19,Yang_PRB21,Mazza_PRA23} in these systems.

\begin{figure*}[t]
        \subfloat[\label{schem}]{\includegraphics[width=.48\textwidth]{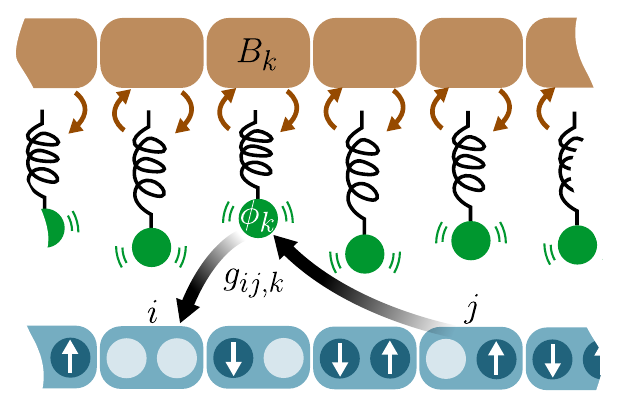}}\quad
        \subfloat[\label{tempfig}]{\includegraphics[width=.48\textwidth]{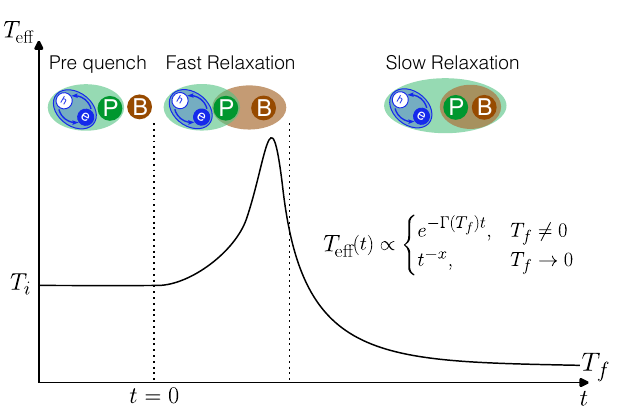}}
        \caption{(a) Schematics of the Yukawa-SYK model where fermions (blue dots with white arrows indicating spin) are scattered by phonons (green dots) from one site to another with a random amplitude $g_{ij,k}$. Each phonon mode $\phi_k$ is coupled to a separate thermal bath $B_k$. (b) Qualitative evolution of the effective temperature in a bath-coupled Yukawa-SYK system. Prior to the quench, phonons and fermion density fluctuations are in a hybrid critical state. Immediately after the quench, the system undergoes a fast process which is followed by a long-lasting process of thermalization where a hydrodynamical description of the system is possible. }
\end{figure*}

While environments are usually perceived as sources of decoherence that adversely affect interesting quantum effects, open quantum systems can host novel phases of matter which are inaccessible in isolated systems \cite{Verstraete_nat2009,dallatorre_nature10,eisert2010noisedriven,KELLY_2021}. Therefore, it is natural to study the dynamics of the SYK model as a prototype for NFL systems in presence of coupling to an environment. The importance of the nature of the bath and the type of system-bath coupling on both the intermediate and long-time behavior of purely fermionic SYK models has been highlighted by Refs. \cite{almheir-syk-2019,zhang-syk-2019,cheipesh2021}. As a controlled theory for the dynamics of a strongly interacting electron-phonon system coupled to an external bath, the Yukawa-SYK model can provide valuable insights into the non-equilibrium dynamics of similar but more complex systems.

We first show that the critical phase survives for couplings to Ohmic, super-Ohmic and a range of sub-Ohmic baths. The distributions of fermionic and phononic excitations display clear qualitative differences over the course of the evolution. Shortly after the quench, the system shows local (in time) equilibrium, where the distribution of excitations is given by a thermal function at an effective, time-dependent temperature. This is similar to the late-time regime of "hydrodynamic" relaxation \cite{Mukerjee_2006,Lux_2014,Bohrdt_2017,Bouchoule_2020,Bastianello_2021} although with important differences which are addressed in Sec. \ref{sec:hydro}. We show that the NFL and Fermi liquid (FL) regimes have different signatures in the thermalization profiles of fermions and phonons. In the NFL regime, fermions thermalize more efficiently than phonons despite the latter's direct coupling to the cold bath. Interestingly, the effective temperature of fermionic excitations is lower than that of phonons at later times. On the other hand, in the FL regime, phonons thermalize faster than fermions and are colder during the intermediate and late stages of thermalization.

The manuscript has the following structure: In Sec. \ref{sec:overview} we demonstrate the theoretical setup and give a summary of our results. In Sec. \ref{sec:model} we introduce the Yukawa-SYK model and characterize different types of baths that we can couple to the system. In Sec. \ref{sec:keldysh} we briefly review the Keldysh formalism and derive the QKE for the Yukawa-SYK model. The results of the numerical solution of QKE and their interpretation are presented in Sec. \ref{sec:results}. Finally, we conclude the paper and propose some future directions in Sec. \ref{sec:conclusions}.

\section{Overview of results}\label{sec:overview}

The Yukawa-SYK model describes a system of Einstein phonons randomly interacting with the density fluctuations of a system of complex fermions (Fig. \ref{schem}). The random fermion-phonon coupling is chosen from a Gaussian ensemble with zero mean and second moment proportional to $g^2$. This system is always in a quantum critical state at low temperatures characterized by the strong hybridization of phonons and fermion density fluctuations \cite{ysyk2019,wang-prl2020,classen-chubukov}. At $t=0$, we couple each phonon species to a separate non-Markovian thermal system of phonons. By assuming a large number of degrees of freedom in each local bath, we neglect the effect of system-bath coupling on the environment. At low energies, the bath density of states (DOS) has a power-law behavior $J(\omega)\sim \omega|\omega|^{a-1}$ where the exponent $a$ characterises the low energy behavior of the environment.

The effect of the bath on the system crucially depends on the scaling behavior of $J(\omega)$ at small frequencies. We call the bath infrared (IR) \textit{irrelevant} when the exponent $a$ given above satisfies $a>a_c$, for a universal value $a_c$ determined only by the Yukawa-SYK model. In this case, the bath DOS goes to zero fast enough at low energies such that the tunneling of phonons between the system and the bath cannot affect the low energy spectrum of the system. The bath is called \textit{marginal} for $a=a_c$, where at low energies, the system-bath coupling and fermion-phonon interactions scale with energy in the same way and similar to the irrelevant bath. Therefore, we do not expect any qualitative change in the spectrum of the system when coupled to a marginal bath. Finally, the bath is termed relevant when $a<a_c$ where we expect the system-bath coupling to affect the low energy spectrum of the system. We will not focus on a relevant bath in the following, since it tends to destroy the critical phase as shown for purely fermionic SYK systems \cite{zhang-syk-2019}. The limitation to an irrelevant bath coupling is not substantial. We will see that it always includes the regime of an Ohmic and super-Ohmic bath and excludes only certain sub-Ohmic environments.

After turning on the system-bath coupling at $t=0$, the evolution of the system follows a two-stage process. An initial and quick stage of dynamics characterized by large energy transfer between the system and the bath as a result of the lack of complete overlap between the eigenstates of pre- and post-quench Hamiltonians followed by a slow and long-lasting period of quasi-equilibrium behavior (Fig. \ref{tempfig}). During the first stage of dynamics, the population of phonons in the system is found to deviate significantly from equilibrium, while the population of fermions appears to be closer to a thermal distribution. As explained below, the relative robustness of the distribution of fermions is a result of Pauli's exclusion principle and the $U(1)$ symmetry of fermions which restrict the available phase space for fermion scattering. The insensitivity of the fermion population allows us to assign fermions an effective temperature using the fluctuation-dissipation ratio (FDR) from the earliest instants of dynamics.

During the second stage of the evolution, both fermions and phonons satisfy fluctuation-dissipation theorem (FDT) at low to moderate energies (with respect to the bare gap of the phonons). The effective temperature can be deduced from FDT (\cite{almheir-syk-2019,zhang-syk-2019}), showing a monotonic decrease toward its final value. Interestingly, fermions appear to be slightly colder than phonons during the second stage of the dynamics, despite the direct coupling of phonons to the bath. We show that the same phenomenon does not occur in a FL variant of the Yukawa-SYK model with the same strength of interactions and conclude that it is a result of the strong correlations between fermions and phonons in the NFL system. We note that such comparison between fermions and phonons is not possible in the quench dynamics of purely fermionic SYK models studied in the past \cite{eberlein,almheir-syk-2019,zhang-syk-2019,cheipesh2021,maldacena-jhep}.

Looking at the evolution of effective temperature quantitatively, we observe an exponential relaxation of temperature and other observables such as energy with the same rate. We call this rate $\Gamma$ which is defined via $T-T_f\sim e^{-\Gamma t}$. We see that $\Gamma$ follows a power-law behavior in terms of the final temperature $\Gamma\sim T_f^x$ where the exponent $x$ is universally determined by the low energy behavior of the system and the bath, and it increases linearly with $a$. This observation suggests that the evolution of every physical quantity is uniquely determined through its dependence on the effective temperature. Thus, the knowledge of the time evolution of the effective temperature is sufficient to describe the behavior of all of the observables. We solve for the complete non-equilibrium dynamics by numerically integrating the QKE. However, to check the mentioned hypothesis and to physically illustrate our numerical results, we  also perform a "hydrodynamic" approximation (see Sec. \ref{sec:hydro} for remarks on the usage of this term) by finding a closed set of equations for the total energy and the rate of energy transfer between the system and the bath as functions of effective temperature. By solving them, we find the dependence of the temperature relaxation rate on final temperature in complete agreement with the results obtained from the numerical integration of QKE. Furthermore, we show that the evolution during the slow phase of the dynamics can itself be separated into two additional stages with different scaling behavior if the bath is at very low temperatures. During the time window where $T\gg T_f$ and for an irrelevant bath, temperature displays power-law behavior in time given by
\begin{equation}
   T_\mathrm{eff}(t)\propto t^{-\frac{1}{a-a_c}},
\end{equation}
while for a marginal bath $T_\mathrm{eff}(t)\propto e^{-\Gamma t}$. For $T-T_f\ll T_f$ there is a crossover where the power-law decay for the irrelevant bath becomes exponential in time. We find the scaling form of the decay rate during this stage to be
\begin{equation}
    \Gamma\propto T_f^{a-a_c},
\end{equation}
for the NFL phase and $\Gamma \propto T^{1+a}$ for FL phase (see Appendix \ref{app:FL}). The faster relaxation at low temperatures can be considered as a fingerprint of the NFL phase.

\section{The model}\label{sec:model}
\subsection{The Yukawa-SYK model}
The Yukawa-SYK model is a system of $N$ phonon and $2N$ fermion species in the exactly solvable limit $N\to \infty$. The Hamiltonian of the system is
\begin{equation}\label{H-ysyk}
    H=\frac12 \sum_{k=1}^{N}\left(\pi_{k}^{2}+\omega_{0}^{2}\phi_{k}^{2}\right)+\frac{1}{N} \sum_{\substack{ijk\\\sigma=\pm}}^N g_{ij,k}\phi_{k}\psi^{\dagger}_{i\sigma}\psi_{j\sigma}.
\end{equation}
The phonon fields $\phi_k$ and their conjugate momenta $\pi_k$ satisfy the commutation relation $\left[ \phi_k,\pi_l\right]=i\delta_{kl}$ and have the bare gap $\omega_0$, while fermion operators are defined by the anti-commutation relation $\{\psi_{i\sigma},\psi^\dagger_{j\sigma'}\}=\delta_{ij}\delta_{\sigma\sigma'}$. The random couplings $g_{ij,k}\equiv g'_{ij,k}+ig''_{ij,k}$ are chosen from Gaussian ensembles with zero means and equal second moments given by
\begin{align}
    \overline{g'_{ij,k}g'_{nm,l}}&=\frac{g^{2}}{2}\left(\delta_{in}\delta_{jm}+\delta_{im}\delta_{jn}\right)\delta_{kl},\\
    \overline{g''_{ij,k}g''_{nm,l}}&=\frac{g^{2}}{2}\left(\delta_{in}\delta_{jm}-\delta_{im}\delta_{jn}\right)\delta_{kl}.
\end{align}
We note that one can define the Yukawa-SYK model for $N$ fermion and $M$ phonon species. The resulting theory is still critical at low temperatures in the limit $N,M\to \infty$ as long as $\frac{N}{M}$ is finite \cite{ysyk2019,classen-chubukov}. The strength of $g'$ and $g''$ can also be different. This will give an effective interaction between fermions in the Cooper channel. Since we are only interested in the dynamics of the model in the normal phase, we increase the pair-breaking and thus tune the superconducting pairing to zero by taking the same variance for $g'$ and $g''$\cite{HAUCK2020}. 

The Hamiltonian in Eq. (\ref{H-ysyk}) becomes exactly solvable in the limit $N\to\infty$ as the vertex corrections vanish in this limit and a closed system of self-consistent equations for fermion and phonon Green's functions can be obtained \cite{ysyk2019}. A key feature of the system is that particle-hole fluctuations always renormalize the effective phonon gap down to zero as $T\to0$. This is a consequence of the large DOS of fermions at low frequencies in the large $N$ limit. Unlike the usual scenario where the quantum critical point is reached by tuning a parameter in the Hamiltonian, this model has self-tuned criticality at sufficiently low temperatures for all values of $g$ and $\omega_0$ \cite{ysyk2019,wang-prl2020,HAUCK2020,classen-chubukov,Wang-prb21,Pan-prr21}. Due to the disappearance of the phonon gap, both fermions and phonons become critical at low temperatures where the system is in the non-Fermi liquid (NFL) phase. In this regime, the diagonal element of imaginary time Green's functions have the following behavior at large time separations
\begin{equation}\label{G_euc}
    G(\tau)\equiv -\frac{1}{2N}\sum_{i\sigma}\langle \psi_{i\sigma}(\tau)\psi^\dagger_{i\sigma}(0)\rangle \propto \frac{\mathrm{sgn}(\tau)}{ |\tau|^{2\Delta}},
\end{equation}
\begin{equation}\label{D_euc}
    D(\tau)\equiv- \frac{1}{N}\sum_k \langle \phi_k(\tau) \phi_k(0) \rangle \propto \frac{1}{|\tau|^{2-4\Delta}}.
\end{equation}
The parameter $\Delta$ is independent of the microscopic details and reads
\begin{equation}\label{Delta}
    \Delta\approx 0.42.
\end{equation}
Therefore, the scaling dimensions of the fields at the NFL fixed point are given by $[\psi]=\Delta$ and $[\phi]=1-2\Delta$. One can show that the effective phonon gap scales with temperature as $\omega_{\mathrm{eff}}^2\sim T^{4\Delta-1}$ at low temperatures. At high temperatures, phonons act as static impurities for fermions and the system is in the universality class of the $\mathrm{SYK}_2$ model given by Eq. (\ref{random_hop}) with constant fermionic DOS at low energies. For more detail about the Yukawa-SYK model, we refer the reader to Refs. \cite{ysyk2019,classen-chubukov}.

\subsection{The Bath}
\subsubsection{General Considerations}
Thermalization is a byproduct of coupling the degrees of freedom in the bath to the degrees of freedom in the system. In general, the Hamiltonian of system-bath coupling is given by
\begin{equation}\label{Hsb_gen}
    H_{\mathrm{SB}}(t)=\sum_{ij}\left( a_{ij} \mathcal{O}^S_i(t)\mathcal{O}^B_j(t) + \mathrm{h.\, c.}\right),
\end{equation}
where the operators $\mathcal{O}^S_i$ and $\mathcal{O}^B_i$ only contain the degrees of freedom in the system and the bath, respectively. In order to proceed, we assume that the effect of the system-bath coupling on the bath degrees of freedom is negligible. This means that, the evolution of $\mathcal{O}^B_i$ is approximately given by $\frac{d}{dt}\mathcal{O}^B_i\approx i[H_B,\mathcal{O}^B_i]$, which is justified as long as the number of degrees of freedom in the bath is much larger than the number of degrees of freedom in the system.

We assume that the temperature of the bath is low enough that the system eventually will end up in the critical phase. As a result, we expect the scaling dimension of $\mathcal{O}^S_i$ to be determined by the scaling dimensions of $\phi_k$ and $\psi_{i\sigma}$ at the NFL fixed point. For $\mathcal{O}^S_i\sim {\psi^\dagger}^n \psi^{n'} \phi^m$ by using Eqs. ($\ref{G_euc}$) and (\ref{D_euc}) we find the scaling dimension of the coupling $a_{ij}$ at the NFL fixed point
\begin{equation}\label{bath_scaling}
    [a_{ij}]=1-(n+n')\Delta-m(1-2\Delta)-[\mathcal{O}^B_j].
\end{equation}
Couplings can be organized according to the sign of their scaling dimension as relevant ($[a_{ij}]>0$), marginal ($[a_{ij}]=0$) and irrelevant ($[a_{ij}]<0$). In the language of renormalization group (RG), the effect of relevant couplings is pronounced at low energies. Marginal and irrelevant couplings are supposed to only alter the non-universal properties of the system, since their effect is suppressed at low energies. This does not mean though, that irrelevant couplings are unimportant as they still can thermalize the system via coupling to high energy modes. 


Based on Eq. (\ref{bath_scaling}), we see that couplings with higher powers of fermion and phonon operators are generally less efficient. Therefore, it is permissible to neglect higher order operators and to only keep those with the lowest order consistent with the symmetries.

Below, we consider two general classes of phonon and fermion baths relevant to the Yukawa-SYK model. Using aforementioned scaling arguments, we will explain that the setup in Fig. \ref{schem} is the physically most relevant one that preserves the NFL phase at low temperatures.

\subsubsection{Phonon Bath}\label{Phononbath}
A phonon bath consists of a set of phonon displacement operators $\{ X_l \}$ together with their conjugate momenta $\{ P_l\}$. We assume that instead of having $H_B$, we know all of the connected correlation functions of the bath. For our purposes, though, only the two-point functions are required
\begin{equation}\label{D_bath_gen}
    \mathfrak{D}_{ij}(t,t')=-i\langle X_{i}(t) X_{j}(t')\rangle.
\end{equation}
We have assumed that the bath is $Z_2$ symmetric, so correlation functions with an odd number of phonon operators vanish. We want to find the net effect of system-bath coupling on the  dynamics of the system. In order to do so, we need to integrate out the bath degrees of freedom. Due to the potentially non-local behavior of the correlation functions in Eq. (\ref{D_bath_gen}), integrating out the bath degrees of freedom can induce correlations in the system that are non-local in time. Also, coupling to the bath can make the off-diagonal Green's functions of the system like $\langle \phi_k \phi_l \rangle$ non-vanishing. While the former effect is actually a desirable feature that can lead to interesting physics, the latter can ruin the exact solvability of the model. Two ways to work around this issue is to use an SYK-like random system-bath coupling where the induced off-diagonal elements vanish in the large $N$ limit \cite{zhang-syk-2019,almheir-syk-2019,cheipesh2021} or to couple each degree of freedom in the system to a separate bath. While we explicitly take the latter route, one can show that the former approach gives similar results.

The lowest order terms in Eq. (\ref{Hsb_gen}) for a phonon bath have the form $X\phi$. Note that the coupling of the bath phonons to density fluctuations of fermions $X\psi^\dagger \psi$ is less relevant than the phonon coupling and can be ignored. We assume that for every phonon mode in the system $\phi_k$, there is effectively one independent but similar mode in the bath $X_k$ coupled to the phonon mode. As a result, we can write
\begin{equation}\label{hsb_phon}
    H_{\mathrm{SB}}=\sum_k \phi_k X_k.
\end{equation}
Therefore, we are taking each $X_k$ to be a separate Caldeira-Leggett bath \cite{caldeira,kamenev}. The bath correlation functions are diagonal and time-translation invariant:
\begin{equation}
    \langle X_{i}(t) X_{j}(t')\rangle=i\delta_{ij}\mathfrak{D}(t-t').
\end{equation}
The response and Keldysh correlation functions of the bath are defined as
\begin{equation}
    \mathfrak{D}^R(t)=\Theta(t)\left[\mathfrak{D}(t)-\mathfrak{D}(-t)\right],
\end{equation}
\begin{equation}
    \mathfrak{D}^K(t)=\mathfrak{D}(t)+\mathfrak{D}(-t).
\end{equation}
The spectral density of phonons in the bath $J(\omega)$ can be found from $J(\omega)=-\mathrm{Im}\, \mathfrak{D}^R(\omega)$. We assume that $J(\omega)$ is given by the generic expression
\begin{equation}\label{ph_bath_spec}
    J(\omega)=\gamma \sin{\left(\frac{\pi a}{2}\right)} \,\mathrm{sgn}\,(\omega) |\omega|^{a} e^{-|\omega|/\omega_c}.
\end{equation}
The parameter $\gamma$ determines the strength of system-bath coupling and $\omega_c$ is a cut-off energy scale assumed to be larger than relevant energy scales in the system. Depending on the value of $a$, the bath is usually called Ohmic ($a=1$), super-Ohmic ($a>1$) and sub-Ohmic ($a<1$). The real part of $\mathfrak{D}^R$ can be found from $J(\omega)$ using Kramers-Kronig relations. The Keldysh function $\mathfrak{D}
^K$ is found from the condition of thermal equilibrium for the bath and using FDT (\ref{FDT}). The low frequency limit of Eq. (\ref{ph_bath_spec}) gives us the scaling dimension of $X$
\begin{equation}
    [X]=\frac{1+a}{2}.
\end{equation}
Putting this in Eq. (\ref{bath_scaling}) we see that $H_{\mathrm{SB}}$ is irrelevant for
\begin{equation}\label{irrel_cond}
a>a_c\equiv 4\Delta-1.
\end{equation}
The value for the threshold exponent $a_c$ coincides with the power of the frequency or temperature divergence of critical phonons\cite{ysyk2019}.  
Hence, the NFL phase of the Yukawa-SYK model survives after coupling the system to a wide range of generic cold phonon baths, unlike coupling to a fermion bath were the simplest system-bath coupling consisting of direct charge transfer between the system and the reservoir destroys the NFL phase as was shown before \cite{almheir-syk-2019,zhang-syk-2019,banarjee2017}. This point becomes relevant by noting that in the real world electron-phonon systems, thermalization dominantly occurs through the interaction of phonons with the environment.

Notice, with the above given value for $\Delta$ follows $a_c\approx 0.68$. Hence, our analysis applies to super-Ohmic, Ohmic, and $a>a_c$ sub-Ohmic baths. By varying the ratio $M/N$ of the phonon and fermion modes the exponent $\Delta$ varies between $\tfrac{1}{4}$ and $\tfrac{1}{2}$\cite{ysyk2019}, such that $0<a_c <1$.  Hence, our analysis is always applicable to the Ohmic and super-Ohmic regime.

\subsubsection{Fermion Bath}
We consider a particle-hole symmetric fermion bath described by the set of fermion operators $(\chi_{i\sigma}^\dagger,\chi_{i\sigma})$. We assume the bath respects $SU(2)$ and $U(1)$ symmetries and therefore, conserves total spin and charge. Again, we use Eq. (\ref{bath_scaling}) and only consider the lowest order terms in $\psi$ and $\phi$ that conserve spin and charge in the system-bath mixture to see that the most relevant terms in $H_{\mathrm{SB}}$ are direct fermion tunneling $\chi^\dagger_{\sigma}\psi_{\sigma}+\psi^\dagger_{\sigma}\chi_{\sigma}$ and fermion-phonon scattering $\chi^\dagger_\sigma \chi_\sigma \phi$. A generic fermion bath has a nearly uniform density of states at low energies and is conveniently modeled by a $\mathrm{SYK}_2$ system defined as
\begin{equation}\label{random_hop}
    H_{\mathrm{SYK2}}=-\frac{1}{\sqrt{N_B}}\sum_{\substack{ij\\ \sigma=\pm}}^{N_B}t_{ij}\chi^\dagger_{i\sigma}\chi_{j\sigma}.
\end{equation}
The random hopping term $t_{ij}=-t_{ij}^\star$ is chosen from a Gaussian ensemble with zero mean and second moment $\overline{|t_{ij}|^2}=t^2$. The Hamiltonian in Eq. (\ref{random_hop}) can be solved after taking the average over the random hopping, resulting in the well-known semi-circular DOS. A convenient way to define the tunneling Hamiltonian such that it preserves the exact solvability of the model is to take
\begin{equation}\label{random_tun}
    H_{\mathrm{SB}}=-\frac{1}{\sqrt{N_B}}\sum_i^N \sum_j^{N_B} \sum_{\sigma=\pm}\left(\alpha_{ij}\psi^\dagger_{i\sigma}\chi_{j\sigma}+\mathrm{h.c.}\right),
\end{equation}
where $\overline{|\alpha_{ij}|^2}=\alpha^2$ and $\overline{\alpha_{ij}}=0$. The pre-factor in Eq. (\ref{random_tun}) ensures that the effect of system-bath coupling on the bath is $\mathcal{O}(N/N_B)$ and therefore, the bath is not affected by $H_{\mathrm{SB}}$ when $N_B\gg N$. According to Eq. (\ref{random_hop}), the scaling dimension of $\chi$ is $\frac{1}{2}$. By putting this in Eq. (\ref{bath_scaling}), we see that (\ref{random_tun}) is a relevant coupling destroying the NFL phase and as a result, the low energy limit of the system is a FL with linearly Landau-damped phonons.

In case of coupling phonons to a bath of fermions described by Eq. (\ref{random_hop}), we can use a Yukawa vertex similar to the original model in Eq. (\ref{H-ysyk})
\begin{equation}
    H_{\mathrm{SB}}=\frac{1}{\sqrt{N N_B}}\sum_k^N \sum_{\substack{ij \\ \sigma=\pm}}^{N_B}\lambda_{ij,k}\chi^\dagger_{i\sigma}\chi_{j\sigma}\phi_k.
\end{equation}
It is easy to check that this coupling is irrelevant and due to the uniform DOS of the bath, contributes to linear Landau damping of phonons similar to second term in Eq. (\ref{rand_hop_p_se_fin}). At low energies, this regime is similar to coupling phonons to an Ohmic phonon bath corresponding to $a=1$ in Eq. (\ref{ph_bath_spec}) and hence, does not require a separate treatment.

\section{Keldysh Formalism}\label{sec:keldysh}
\subsection{General Definitions}
\begin{figure}[t]
    \centering
    \includegraphics[width=0.48\textwidth]{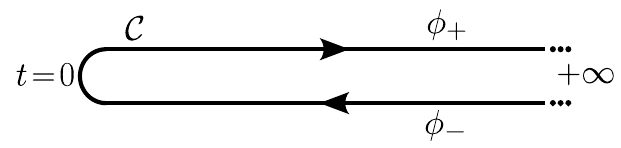}
    \caption{The closed time contour in the Schwinger-Keldysh formalism starts at the initial moment of the evolution ($t=0$) and goes to infinite future and back. Every field is defined on $\mathcal{C}$. Equivalently, each field is decomposed to its forward and backward components.}
    \label{keldysh_cont}
\end{figure}
We only mention briefly the essential concepts used in our work and refer readers to Ref. \cite{kamenev} for a detailed treatment of Keldysh formalism. In Keldysh approach, we work with greater $G^>$ and lesser $G^<$ correlation functions
\begin{align}\label{green_gl}
    G^>(t,t')&\equiv-i\langle \hat{a}(t)\hat{a}^\dagger(t')\rangle, \\
    G^<(t,t')&\equiv-i\xi\langle \hat{a}^\dagger(t')\hat{a}(t)\rangle,
\end{align}
where $\xi=\pm$ corresponds to bosonic ($+$) and fermionic ($-$) statistics for $\hat{a}$. In the equilibrium formalism, the central objects of study are time-ordered correlation functions and the \textit{physically measurable} correlation functions are only found at the end of calculation using analytical continuation whereas in Keldysh field theory the response and Keldysh (symmetric) correlation functions can be found directly from Eq. (\ref{green_gl})
\begin{align}
    G^R(t,t')&\equiv\Theta(t-t')\left(G^>(t,t')-G^<(t,t')\right), \label{ret_def} \\
    G^A(t,t')&=\left[G^R(t,t')\right]^\dagger, \label{adv_def} \\
    G^K(t,t')&\equiv G^>(t,t')+G^<(t,t'). \label{keld_def}
\end{align}
In Keldysh field theory we are not limited to thermal states and in general, the evolution of Green's functions in (\ref{green_gl}) is given by QKE which are a set of self-consistent integro-differential equations between correlation functions of different order. For a generic interacting system there are an infinite number of these equations, a quantum counterpart of the BBGKY hierarchy \cite{kardar_2007}. Accordingly, a truncation of QKE is often required which inevitably results in approximate solutions. For SYK models however and as we show below, QKE become closed at the level of 2-point functions in the limit $N\to \infty$, allowing us to monitor the dynamics exactly.

For steady states, Green's functions depend on $t-t'$, allowing us to take their Fourier transform. The spectral density of single-particle states $\mathcal{A}(\omega)$ is determined by
\begin{equation}\label{spectral_func}
    \mathcal{A}(\omega)=-2\,\mathrm{Im}\, G^R(\omega). 
\end{equation}
Particularly in thermal equilibrium and at temperature $T$, the fluctuation-dissipation theorem (FDT) relates $G^K(\omega)$ to $\mathcal{A}(\omega)$ via
\begin{equation}\label{FDT}
    \left( \frac{iG^K(\omega)}{\mathcal{A}(\omega)}\right)^\xi=\tanh{\frac{\omega}{2T}}.
\end{equation}
One usually can regard Eq. (\ref{FDT}) as a relation that gives $G^K$ in terms $\mathcal{A}$ when the temperature is known. Out of equilibrium, time-translation symmetry is usually broken and Green' functions are not only functions of $t-t'$. As a result, there is no a unique way to extend Eqs. (\ref{spectral_func}) and (\ref{FDT}) to out-of-equilibrium situations. Two commonly used extensions are the Wigner transformation defined by \cite{kamenev} 
\begin{equation}\label{wigner}
    G(t,\omega)\equiv \int_{-\infty}^{+\infty}G(t+\frac{\tau}{2},t-\frac{\tau}{2}) e^{i\omega \tau}\,d\tau,
\end{equation}
and Fourier transformation along the "corner slice" given by \cite{maldacena-jhep}
\begin{multline}\label{corner}
    G(t,\omega)\equiv \int_{-\infty}^{+\infty}\left[ \Theta(\tau) G(t,t-\tau) \right. \\ \left. +\Theta(-\tau) G(t+\tau,t) \right] e^{i\omega \tau}\,d\tau.
\end{multline}
Both definitions are expected be equivalent for steady states and also non-steady states if the change in $G(t,t')$ along the center of mass coordinate $\frac{t+t'}{2}$ is slower than the change along $t-t'$. As a result, the two definitions in Eqs. (\ref{wigner}) and (\ref{corner}) can give quite different results out of equilibrium. Although the Wigner transformation is the most commonly used definition, it violates the causal structure of kinetic equations given in Eq. (\ref{causal_cond}) below. The same does not happen for integration over the corner slice which is actually the natural choice considering how Green's functions are evolved in time by QKE (Appendix. \ref{appen_kinetic}). For instance, in a quench at $t=t_0$, the function $G(t,\omega)$ defined according to Eq. (\ref{wigner}) displays non-trivial dynamics at $t<t_0$, before the quench happens. This issue becomes more pronounced in critical systems where memory effects are strong due to the slow decay of correlations in time. Henceforth, we employ the corner slice to define time-dependent Green's functions in the frequency domain. The spectral density at time $t$ is defined analogously to Eq. (\ref{spectral_func}). We define the fluctuation-dissipation ratio (FDR) as
\begin{equation}\label{FDR}
    F(t,\omega)= \left( \frac{iG^K(t,\omega)}{\mathcal{A}(t,\omega)}\right)^\xi,
\end{equation}
which can be used to find the non-equilibrium one-particle distribution function $n(t,\omega)$ according to \cite{kamenev}
\begin{equation}
    F(t,\omega)=\qty(1+ 2 \xi \, n(t,\omega))^{-\xi}.
\end{equation}
An effective temperature $T_\mathrm{eff}(t)$ can be defined if for small frequencies $n(t,\omega)$ is approximated by the Bose-Einstein (Fermi-Dirac) distribution for bosons (fermions)
\begin{equation}\label{eff_T}
    \tanh{\frac{\omega}{2T_\mathrm{eff}(t)}}\approx F(t,\omega).
\end{equation}
Note that, in out-of-equilibrium situations the temperature is found by the best fitting of a hyperbolic function to $F(t,\omega)$. In equilibrium, FDT is used to obtain $G^K$ in terms of spectral density and temperature.

Keldysh field theory can also be expressed in the path integral language. This can be easily seen by looking at the evolution of density matrix
\begin{equation}
    \rho(t)=U(t,0)\rho_0 U^\dagger(t,0),
\end{equation}
where $\rho_0$ is the initial density matrix and $U$ is the unitary time evolution operator. By decomposing time evolution operators on each side into the multiplication of time evolution operators over small time steps, inserting resolutions of identity between them and applying a Trotter expansion, we get a path integral for each side corresponding to forward (left side) and backward (right side) directions of integration. The quantum fields of opposite directions are not independent. They are coupled through the matrix element of $\rho_0$ between the fields of opposite contours at $t=0$. Moreover, since the trace of $\rho(t)$ is to be calculated eventually, the value of the fields on opposite branches should coincide at the final time. As a result, one can define quantum fields on a closed time contour $\mathcal{C}$ shown in Fig. \ref{keldysh_cont}. Naively, the usual temporal integration in the action is replaced by integration over the contour
\begin{equation}
    \int_{-\infty}^{+\infty} L(t)\, dt \to \oint_{\mathcal{C}} L(t_c) \, dt_c,
\end{equation}
while keeping in mind to take care of the boundary conditions mentioned above. Instead of time ordering in the equilibrium formalism, the correlation functions are contour ordered
\begin{equation}
    iG(t_c,t'_c)\equiv \langle T_{\mathcal{C}}a(t_c)\bar{a}(t'_c)\rangle=\xi \langle T_{\mathcal{C}}\bar{a}(t'_c)a(t_c)\rangle,
\end{equation}
where $T_{\mathcal{C}}$ is the contour ordering operator.
Equivalently, we can assign each field an extra index corresponding to whether it is on the forward or backward branches of the contour. Then, we have an alternative expression for the greater and lesser correlation functions given in Eq. (\ref{green_gl}) as
\begin{align}\label{green_def}
    G^>(t,t')&\equiv G(t-,t'+), \\
    G^<(t,t')&\equiv G(t+,t'-).
\end{align}
\subsection{Keldysh Action for the Yukawa-SYK Model}
\begin{figure}[t]
    \subfloat[\label{int_ver}]{\includegraphics[scale=0.5]{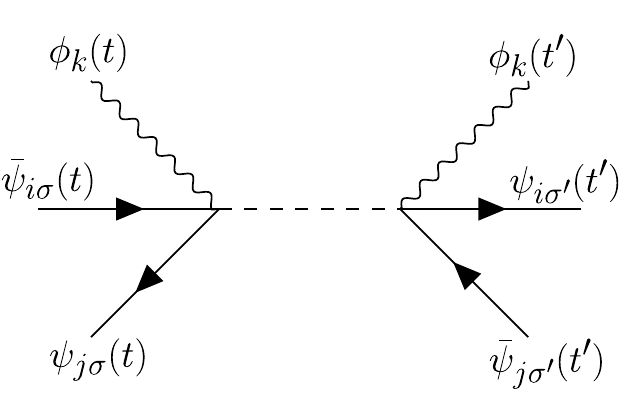}} \\
    \subfloat[\label{ferm-dyson}]{\includegraphics[scale=0.9]{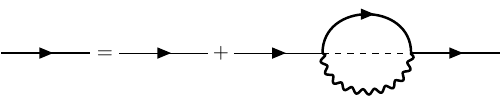}}\\
    \subfloat[\label{phon-dyson}]{
    \includegraphics[scale=0.9]{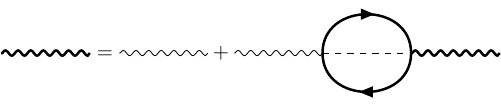}}
    \caption{(a) The interaction vertex for the Yukawa-SYK model after averaging over random interactions. (b) and (c) Diagrammatic representation of Dyson equations for (b) fermions and (c) phonons. Narrow lines are bare Green's functions.}
\end{figure}

When the system is isolated, its evolution follows Eq. (\ref{H-ysyk}). Therefore, we can use the Keldysh action of Eq. (\ref{H-ysyk}) to describes the dynamics. This action is given by
\begin{equation}
S=S_{\mathrm{f}}+S_{\mathrm{ph}}+S_{\mathrm{int}},
\end{equation}
\begin{equation}\label{s_f}
    S_{\mathrm{f}}=\sum_{\substack{i\\ \sigma=\pm}}^{N} \oint_{\mathcal{C}} dt_{c} \, i\bar{\psi}_{i\sigma}(t_{c})\partial_{t_{c}}\psi_{i\sigma}(t_{c}),
\end{equation}
\begin{equation}\label{s_ph}
    S_{\mathrm{ph}}=-\frac12 \sum_{k}^{N} \oint_{\mathcal{C}}dt_{c}\,  \left(\dot{\phi}_k^2+\omega_{0}^{2}\phi_k^2\right),
\end{equation}
\begin{equation}\label{s_int}
    S_{\mathrm{int}}=-\frac{1}{N} \sum_{\substack{ijk\\ \sigma=\pm}}^N g_{ij,k} \oint_{\mathcal{C}}dt_{c}\, \phi_{k}(t_{c})\bar{\psi}_{i\sigma}(t_{c})\psi_{j\sigma}(t_{c}).
\end{equation}
In Keldysh formalism, disorder averaging is implemented directly, without the need to use methods like replica trick \cite{Edwards_1975} or supersymmetry \cite{efetov_1996}. By evaluating the Gaussian integrals over disorder realizations, we get the effective interaction vertex in Fig. \ref{int_ver} given by the non-local action
\begin{multline}\label{s_eff}
    S_{\mathrm{eff}}=\frac{ig^{2}}{2N^2} \sum_{ij,k}^N\sum_{\sigma, \sigma'=\pm} \oint\oint  dt_{c}dt_{c}' \, \phi_{k}(t_{c})\phi_{k}(t'_{c}) \\ \times \bar{\psi}_{i\sigma}(t_{c})\psi_{j\sigma}(t_{c})\bar{\psi}_{j\sigma'}(t'_{c})\psi_{i\sigma'}(t'_{c}).
\end{multline}
The fermion and phonon Green's functions are defined by
\begin{align}
  G(t_c,t'_c)&=-i\langle T_{\mathcal{C}} \psi_{i\sigma}(t_c) \bar{\psi}_{i\sigma}(t_c')\rangle, \\
  D(t_c,t'_c)&=-i\langle T_{\mathcal{C}} \phi_k(t_c) \phi_k(t_c')\rangle, \label{D(t)}
\end{align}
and are assumed to be independent of field indices. The Green's functions satisfy Dyson equations
\begin{equation}\label{dyson_f_compact}
    G=G_0+G_0\otimes \Sigma \otimes G,
\end{equation}
\begin{equation}\label{dyson_ph_compact}
    D=D_0+D_0\otimes \Pi \otimes D.
\end{equation}
 $G_0$ and $D_0$ are fermion and phonon Green's functions in the absence of interactions and $\Sigma$ and $\Pi$ are fermion and phonon self-energies, respectively. For $N\to \infty$, vertex corrections can safely be ignored and self-energies are given by loop diagrams in Figs. \ref{ferm-dyson} and \ref{phon-dyson} which read as
\begin{align}
    \Sigma(t_c,t'_c)&=ig^2 G(t_c,t'_c)D(t_c,t'_c),\label{f_self} \\
    \Pi(t_c,t'_c)&=-2ig^2 G(t_c,t'_c)G(t'_c,t_c). \label{ph_self}
\end{align}
It may appear that the Dyson equations should be solved self-consistently. Nevertheless, by applying the inverse of free Green's functions on both sides of Eqs. (\ref{dyson_f_compact}) and (\ref{dyson_ph_compact}) we get
\begin{equation}\label{dyson_f}
    i\partial_{t_{c}}G(t_{c},t'_{c})=\delta(t_{c},t'_{c}) + \oint \Sigma(t_{c},\tau_{c})G(\tau_{c},t'_{c})\, d\tau_{c},
\end{equation}
\begin{multline}\label{dyson_ph}
    -\left( \partial_{t^{ }_{c}}^{2}+\omega_{0}^{2}\right)D(t_{c},t'_{c})=\delta(t_{c},t'_{c})\\+ \oint \Pi(t_{c},\tau_{c})D(\tau_{c},t'_{c})\, d\tau_{c}.
\end{multline}
One can show that (see Appendix \ref{appen_kinetic}) these equations have a causal structure such that the value of a function at $(t,t')$ only depends on the value of other functions at times $(t_1,t_2)$ satisfying
\begin{equation}\label{causal_cond}
    \max \{t_1,t_2\}<\max\{t,t'\},
\end{equation}
and therefore, we do not need to solve Dyson equations self-consistently. In order to find the dynamics, we have to write Eqs. (\ref{dyson_f}) and (\ref{dyson_ph}) in terms of greater and lesser functions and then solve them numerically. While these equations together with FDT are sufficient to find the Green's functions at equilibrium, an accurate numerical solution of integro-differential equations for time evolution requires us to work with first order time derivatives. Therefore, we rewrite the equations for phonons (\ref{dyson_ph}) in terms of a larger set of first order equations. This is achieved by introducing two extra correlation functions
\begin{align}
    B(t_c,t'_c)&=-i\langle T_{\mathcal{C}}\pi_k(t_c) \phi_k(t'_c) \rangle, \label{B(t)}\\
    C(t_c,t'_c)&=-i\langle T_{\mathcal{C}}\pi_k(t_c) \pi_k(t'_c) \rangle, \label{C(t)}
\end{align}
where $\pi_k=\dot{\phi}_k$ is the momentum conjugate field of $\phi_k$. A detailed discussion of the first order quantum kinetic equations and their numerical solution can be found in Appendix \ref{appen_kinetic}.

\subsection{Formulating System-Bath Coupling}
Here, we only consider the coupling of phonons to a phonon bath as represented by Eq. (\ref{hsb_phon}). For a discussion of fermion bath with the coupling in Eq. (\ref{random_tun}) in Keldysh language, see Ref. \cite{cheipesh2021}. 

Following the arguments of Sec. \ref{Phononbath}, we assume a quadratic Keldysh action for the phonon bath
\begin{equation}\label{s_b_ph}
    S_{\mathrm{B}}=\frac12 \sum_k^N \oint \oint X_k(t_c) \mathfrak{D}^{-1}(t_c,t'_c) X_k(t'_c)\, dt_c\, dt'_c.
\end{equation}
The bath Green's function $\mathfrak{D}$ is determined by Eq. (\ref{ph_bath_spec}) and using FDT. Integrating out the bath degrees of freedom and using Eq. (\ref{hsb_phon}) gives the action responsible for the thermalization of phonons
\begin{equation}\label{s_sb_phon}
    S_{\mathrm{SB}}=-\frac12 \sum_k^N \oint \oint \phi_k(t_c) \mathfrak{D}(t_c,t'_c) \phi_k(t'_c)\, dt_c\, dt'_c,
\end{equation}
which contributes to phonon self-energy in Eq. (\ref{ph_self}) 
\begin{equation}\label{total_ph_self}
    \Pi(t_c,t'_c)=-2ig^2G(t_c,t'_c)G(t'_c,t_c)+\mathfrak{D}(t_c,t'_c).
\end{equation}

\section{Results}\label{sec:results}
Below, we present the results of the numerical solution of QKE (Eqs. (\ref{dyson_f}) and (\ref{dyson_ph})) in Sec. \ref{sec:two_stages} and by looking at the behavior of effective temperature and the deviation of the system from equilibrium at low energies, we motivate a two-stage picture for dynamics after the quench. In Sec. \ref{sec:first_stage} we analyze the first stage of dynamics detail and address the role of phonons and symmetries in the dynamics at early times. The evolution of the system during the second stage of dynamics together with an analytical evaluation of the behavior of relaxation rate are given in sections \ref{sec:second_stage} and \ref{sec:hydro}, respectively.

\subsection{The two-stage picture of post-quench evolution}\label{sec:two_stages}
\begin{figure}[t]
    \centering
    \includegraphics[width=0.48\textwidth]{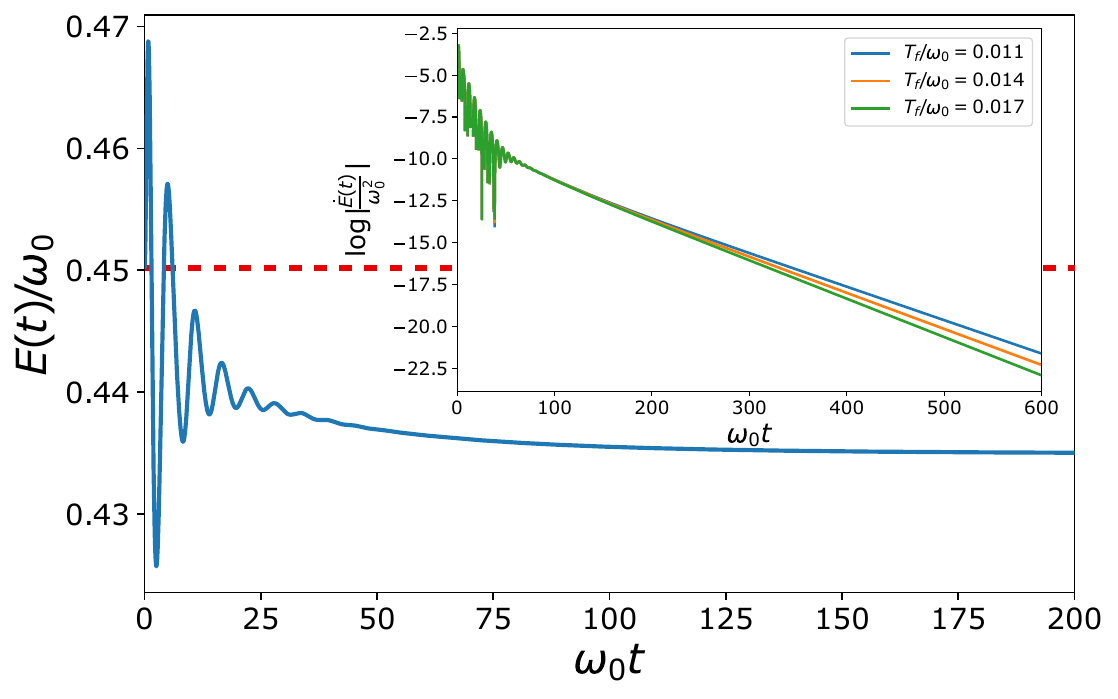}
    \caption{Total energy as a function of time for $g^2/\omega_0^3=0.7$, $a=1.0$, $T_f/\omega_0=0.012$ and $\gamma/\omega_0=0.12$. The red dashed line indicates the total energy before the quench. Inset shows the exponential decay of energy at later times with a relaxation rate (given by the slope of the logarithm) that increases with $T_f$.}
    \label{fig:E(t)}
\end{figure}
\begin{figure}[t]
    \centering
    \subfloat[\label{fig:Ph_DOS}]{
    \includegraphics[width=0.48\linewidth]{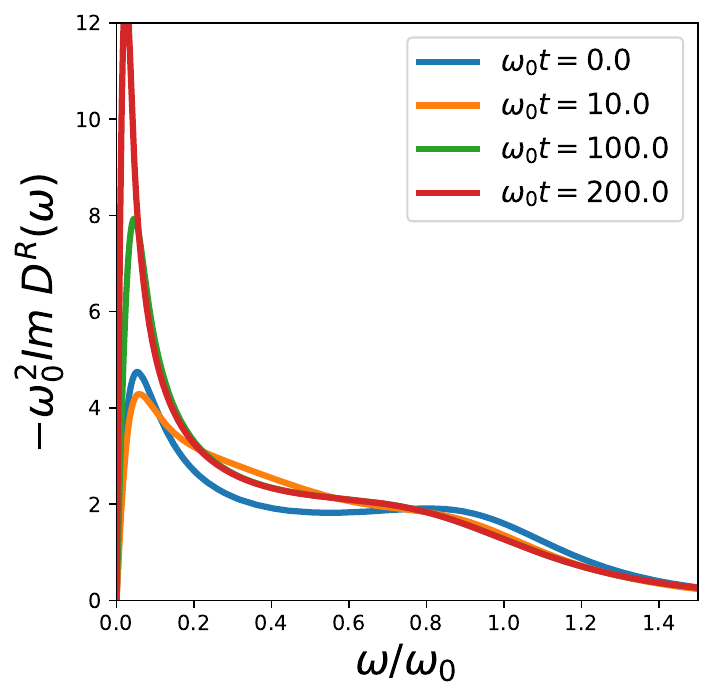}}
    \subfloat[\label{fig:F_DOS}]{
    \includegraphics[width=0.48\linewidth]{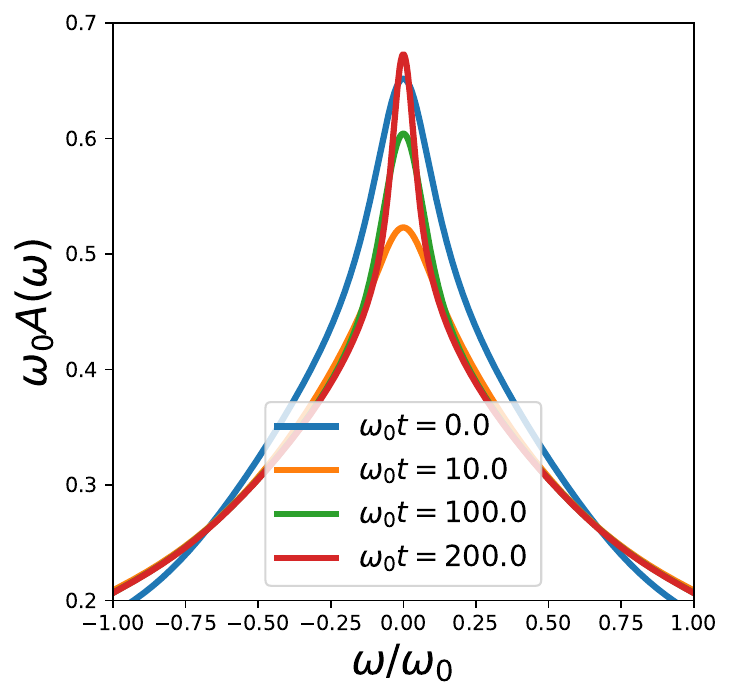}}\\
    \subfloat[\label{fig:Ph_FDR}]{
    \includegraphics[width=0.48\linewidth]{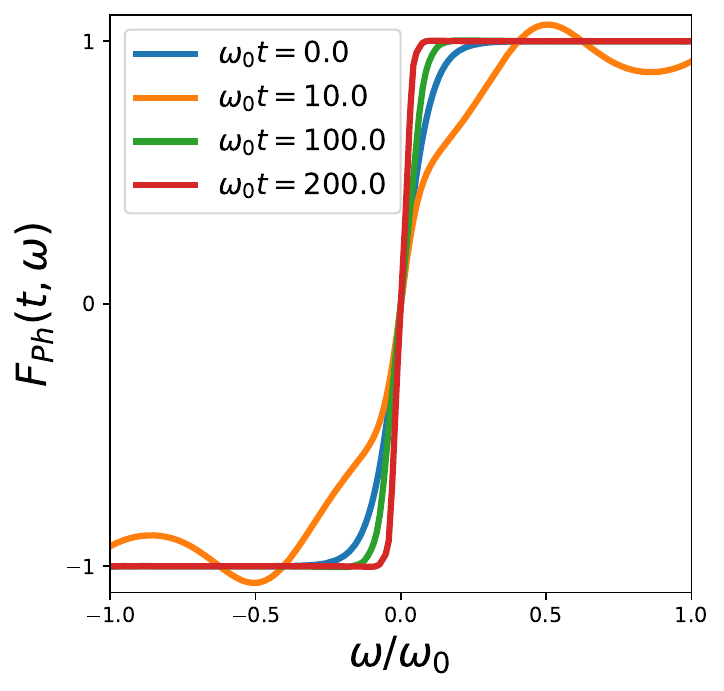}}
    \subfloat[\label{fig:F_FDR}]{
    \includegraphics[width=0.48\linewidth]{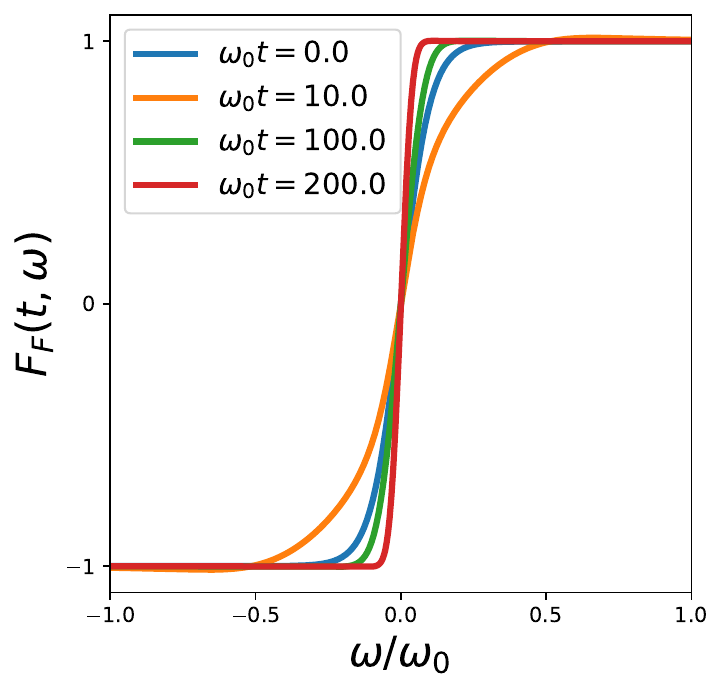}}
    \caption{Top panel: phonon (a) and fermion (b) spectral densities at different times. We can see a quick deformation in both spectral densities at early times. Bottom panel: Fluctuation-dissipation ratio (FDR) for phonons (c) and fermions (d) as a function of frequency at different times. The large early deviation of $F(t,\omega)$ from quasi-equilibrium for phonons is clear. After a while, both fermions and phonons display quasi-equilibrium behavior. The quench parameters are chosen as $g^2/\omega_0^3=0.7$, $a=1.0$, $\gamma/\omega_0=0.12$, $T_i/\omega_0=0.05$ and $T_f/\omega_0=0.012$.}
\end{figure}
\begin{figure}[t]
    \centering
    \includegraphics[width=0.48\textwidth]{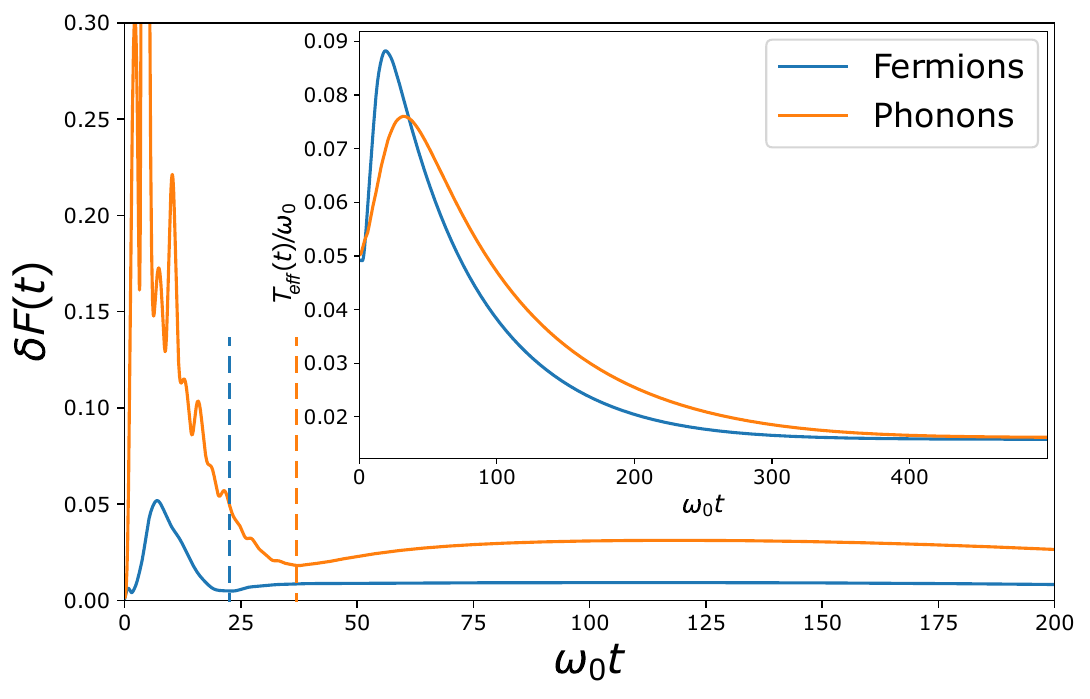}
    \caption{The deviation of FDR for fermions and phonons from quasi-equilibrium defined in Eq. (\ref{delta_F}) as a function of time. The minima show the transition to the second stage of the dynamics. Furthermore, phonons exhibit considerably larger deviations compared to fermions at all times. Dashed lines indicate the the minima of $\delta F$. Inset: The extrapolated effective temperature for fermions and phonons as a function of time. The peaks approximately mark the onset of the second stage of the evolution. The parameters are $g^2/\omega_0^3=0.7$, $a=1.0$, $T_i/\omega_0=0.05$ and $\gamma/\omega_0=0.12$.}
    \label{fig:FDR_T}
\end{figure}
 In this part, we demonstrate the qualitative difference between the behavior of various physical quantities at the early and later periods of the post-quench evolution. Accordingly, we separate the dynamics into two stages as was mentioned in Sec. \ref{sec:overview}. By inspecting the total energy (see Appendix \ref{app:total_energy}) as a function of time shown in Fig. \ref{fig:E(t)}, we observe an oscillatory behavior during the early times after the quench which shortly afterwards, turns into a monotonic decrease until the system reaches its final thermal state at $t\to \infty$. This change in the behavior of the energy is the first sign of the two-stage time evolution of the system.

The one particle spectra of fermions and phonons are depicted in Figs. \ref{fig:Ph_DOS} and \ref{fig:F_DOS} at different times. A quick deformation of both fermionic and phononic spectra immediately after the quench is observed throughout the frequency space. On the other hand, the variation in the spectra is slow and gradual at later stages and mostly occurs at low energies. This crossover in the behavior of the evolution of one particle spectrum from early to later stages of the evolution is another manifestation of the fact that the early and later periods require separate physical descriptions.

Another quantity of interest is the distribution of single particle excitations related to the FDR defined in Eq. (\ref{FDR}). As it can be seen in Figs. \ref{fig:Ph_FDR} and \ref{fig:F_FDR}, the distributions of both phonons and fermions deviate from quasi-equilibrium at early times while at later times they appear to be quite close to a thermal state with a time-dependent temperature. A quantitative measure of the deviations from quasi-equilibrium can be defined if we assign each species an effective temperature which is read from the slope of the FDR at the origin of the frequency space according to:
\begin{equation}\label{T_eff}
    T_{\mathrm{eff}}(t)\approx \frac{1}{2\partial_\omega F(t,\omega)}\bigg|_{\omega\to 0}.
\end{equation}
The effective temperatures obtained from Eq. (\ref{T_eff}) for fermions and phonons is shown in Fig. \ref{fig:FDR_T}. With the exception of the early times, phonons appear to be at a higher temperature than fermions. This may look counter-intuitive from a classical point of view, as phonons are directly coupled to the cold bath while fermions exchange heat with the bath only indirectly via phonons. Nonetheless, quantum effects can explain this behavior as the correlations between fermions and phonons due to their strong interactions render the aforementioned distinction meaningless. This behavior of the strongly interacting NFL phase can be contrasted to the behavior of a similar system with an additional moderate to strong random hopping term for fermions which makes the system a Fermi liquid. In the latter case (see Appendix. \ref{app:FL}), the effective temperature of phonons is smaller than the effective temperature of fermions during the intermediate and later stages of the evolution as the correlations between fermions and phonons are weak compared to the NFL phase.

We use the effective temperatures to define the quasi-equilibrium FDRs for phonons and fermions according to
\begin{equation}\label{F_eq}
    F_{\mathrm{eq}}(t,\omega)\equiv \tanh{\frac{\omega}{2T_{\mathrm{eff}}(t)}}.
\end{equation}
The deviation from quasi-equilibrium can be defined \cite{zhang-syk-2019} in terms of the functional norm of the difference between $F(t,\omega)$ found from Eq. (\ref{FDR}) and $F_{\mathrm{eq}}$ given by Eq. (\ref{F_eq})
\begin{equation}\label{delta_F}
    \delta F(t)\equiv \left[\frac{\int_{-\Lambda}^{+\Lambda} \left[F(t,\omega)-F_{\mathrm{eq}}(t,\omega) \right]^2\,d\omega}{\int_{-\Lambda}^{+\Lambda} F_{\mathrm{eq}}^2(t,\omega)\,d\omega} \right]^{\frac12}.
\end{equation}
We have divided the difference by the norm of $F_{\mathrm{eq}}$ to obtain the relative deviation. The relative deviation $\delta F(t)$ is shown in Fig. \ref{fig:FDR_T} for phonons and fermions. We observe that for both phonons and fermions, there is a temporary increase in $\delta F(t)$ followed by a minimum and then a gradual approach towards the true equilibrium at later times. The oscillations in $E(t)$ stop approximately around the same time (Fig. \ref{fig:E(t)}) when $\delta F(t)$ for fermions and phonons reach their minimum. Accordingly, the minima in the deviations of fermions and phonons from quasi-equilibrium set a natural boundary between the first and second stages of the evolution. 

The separation of the evolution into two stages can also be observed by looking at the behavior of the effective temperature for fermions. We read the effective temperature from the FDR for fermions as their deviation from quasi-equilibrium is significantly less than phonons throughout the entire evolution. As it can be seen in Fig. \ref{fig:FDR_T}, there is a temporary increase in the effective temperature after the quench and as expected from the two-stage picture given above, the location of the peak in the effective temperature is close to the location of the minimum in $\delta F(t)$ for fermions. 

Having explained the qualitative differences between early stage and late stage dynamics, we will separately discuss their properties in more details in the following sections.

\subsection{First stage of dynamics}\label{sec:first_stage}
The first distinct feature of the first stage of dynamics is the fast relaxation of both fermion and phonon densities of states (Fig. \ref{fig:Ph_DOS} and \ref{fig:F_DOS}). Similar behavior has been found in the far from equilibrium dynamics of purely fermionic SYK models \cite{schiro-prb22}. This behavior is pronounced at higher frequencies where the high frequency components of spectral functions at early times coincide with their value at $t\to \infty$ (see the curves for $\omega_0 t=10.0$ in Figs. \ref{fig:Ph_DOS} and \ref{fig:F_DOS}).
 
 The second feature is the early oscillatory evolution of total energy. This behavior has not been observed in the thermalization of purely fermionic SYK models \cite{almheir-syk-2019,zhang-syk-2019} coupled to IR irrelevant external baths. The appearance of these oscillations may seem inconsistent with the Yukawa-SYK model being in a scale invariant critical state with no characteristic energy scales besides the temperature itself. However, the coupling of the system to the bath introduces new energy scales including the system-bath coupling and the UV cut-off $\omega_c$ defined in Eq. (\ref{ph_bath_spec}) which puts a soft upper bound on the energy spectrum of the environment. Furthermore, due to the temporary heating of the system as a result of  the sudden coupling of the system to the bath, the system is pushed away from the critical state. This can render the dynamics sensitive to the bare phonon gap $\omega_0$ and the coupling $g^2$. Phonons as harmonic oscillators exhibit oscillatory behavior and as it can be seen in Figs. \ref{fig:Ph_DOS} and \ref{fig:Ph_FDR}, high frequency phonons with finite spectral weight are generated in the range of frequencies $\omega \lesssim \omega_0$ after the quench, resulting in the emergence temporary oscillations in the profile of phonon correlation functions over time scales $t\gtrsim 1/\omega_0$ in agreement with the approximate period of initial energy oscillations $\tau\approx 2\pi/\omega_0$ in Fig. \ref{fig:E(t)}. For the Yukawa-SYK model, the total energy is given by 
  \begin{equation}\label{simple_energy}
     \frac{E(t)}{N}=\langle \pi^2 \rangle,
 \end{equation}
where the contribution of the potential term $(\sim \frac12 \omega_0^2 \phi^2)$ is cancelled by the fermion-phonon interaction while the kinetic term is amplified by a factor of 2. We refer the reader to Appendix \ref{app:total_energy} for a derivation of this result. The above expression holds in and out of equilibrium and it directly connects the energy oscillations to the oscillations of phonon correlator as explained before.
 
The third distinct feature of the first stage of dynamics is the deviation of the populations of both fermions and phonons from a thermal distribution (Figs. \ref{fig:Ph_FDR} and \ref{fig:F_FDR}). While a temporary distortion in the distribution of particles is naturally expected as a result of the sudden coupling of the system to the bath, the relative robustness of the distribution of fermions compared to phonons as it can be seen in Fig. \ref{fig:FDR_T} requires extra physical explanation. At first, one may argue that the substantial difference between the distribution of phonons and fermions comes from coupling the bath directly to phonons while fermions are affected less as they only indirectly interact with the bath through their mutual coupling to phonons. However, this statement cannot fully explain the physics of the problem for two reasons: First, we are dealing with a moderately strong interacting system of fermions and phonons and therefore, we expect any perturbations in the phononic sector to be transmitted efficiently to the fermionic sector. Second, the rigidity of the distribution of fermions persists even in case of directly coupling fermions to the bath as long as one crucial condition (see below) is satisfied. The robustness of fermionic distribution can be attributed to two elements: the Pauli's exclusion principle due to fermionic statistics and the global $U(1)$ symmetry of the problem under the transformation $\psi_{i\sigma} \to e^{i\varphi}\psi_{i\sigma}$ which guarantees fermion number conservation. To the extent of our knowledge, the role of $U(1)$ symmetry in the quench dynamics of SYK models has not been investigated before. We will explain below, using Fermi's golden rule arguments, how $U(1)$ symmetry and Fermi statistics restrict the distortion of the fermionic distribution function after a quench.
 
 When we turn on the system-bath coupling by a quench function $f(t)$ (for this work $f(t)\propto \Theta(t)$), excitations are created in the system (the bath is large and assumed to be unaffected by the quench). The energy of these excitations depends crucially on the power spectrum of $f(t)$ defined as $|\Tilde{f}(\omega)|^2$ where $\Tilde{f}(\omega)$ is the Fourier transform of $f(t)$. The perturbative rate of change in the phonon distribution at energy $\omega$, consistent with the symmetries of the problem to the lowest order in the system bath coupling satisfies
 \begin{multline}\label{nph_dot}
     \partial_t n_{\mathrm{ph}}(\omega)\propto \int |\tilde{f}(\epsilon)|^2 J(\omega-\epsilon) \, \rho_{\mathrm{ph}}(\omega)\\ \times \left(n_{\mathrm{B}}(\omega-\epsilon)-n_{\mathrm{ph}}(\omega)\right)\, d\epsilon,
 \end{multline}
where $\rho_{\mathrm{ph}}$ is the phonon spectral density, $n_{\mathrm{B}}$ is the distribution of phonons in the bath and $J(\omega)$ is the bath spectral density given in Eq. (\ref{ph_bath_spec}). For a direct coupling between fermions and a phononic bath respecting the $U(1)$ symmetry we have
\begin{multline}\label{nf_dot}
    \partial_t n_{\mathrm{f}}(\omega)\propto \int d\epsilon \int d\nu \, |\Tilde{f}(\epsilon)|^2 \mathcal{A}(\omega)J(\nu)(n_{\mathrm{B}}(\nu)-1) \\ \times \left\{(1-n_{\mathrm{f}}(\omega))n_{\mathrm{f}}(\omega-\epsilon-\nu) \mathcal{A}(\omega-\epsilon-\nu) \right. \\ \left. - n_{\mathrm{f}}(\omega)(1-n_{\mathrm{f}}(\omega+\epsilon+\nu))\mathcal{A}(\omega+\epsilon+\nu)\right\} \\ + \, \text{higher order terms with even powers of}\, \, \mathcal{A}(\omega),
\end{multline}
where $\mathcal{A}(\omega)$ is the fermion spectral density. The fact that we always have even powers of $\mathcal{A}$ is a consequence of the $U(1)$ symmetry in the system. Together with fermionic statistics, this severely restricts the contributing domain of integration in Eq. (\ref{nf_dot}) to excitations close to the Fermi energy. This is in contrast to Eq. (\ref{nph_dot}) for phonons where such restrictions do not hold.

\subsection{Second stage of dynamics}\label{sec:second_stage}
\begin{figure}[t]
    \centering
    \includegraphics[width=0.48\textwidth]{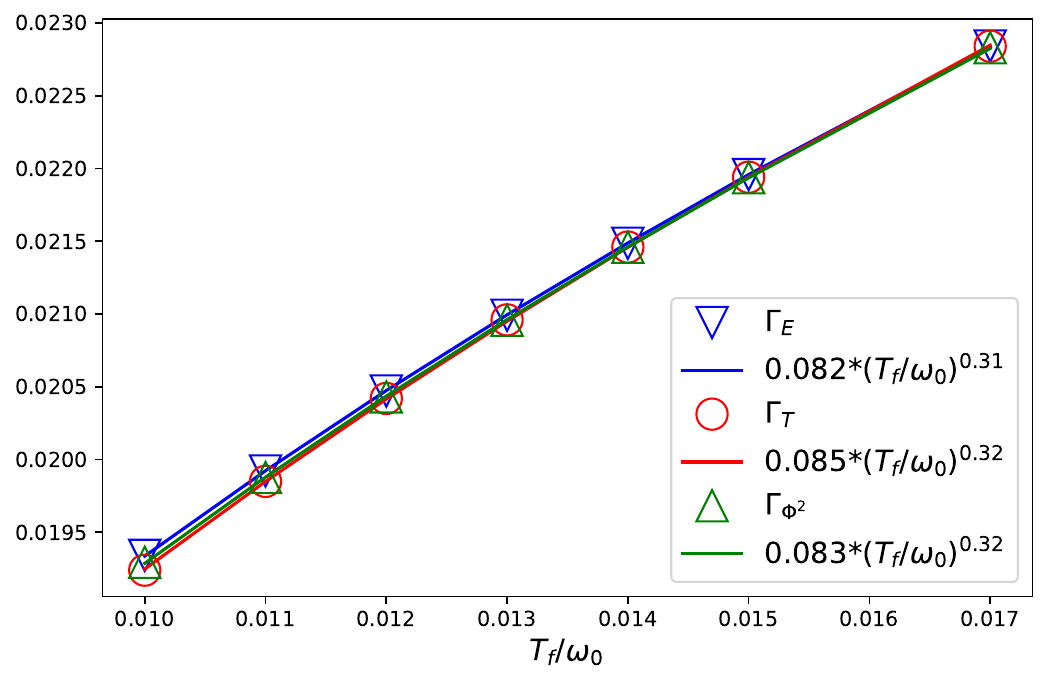}
    \caption{The relaxation rates of temperature, energy and $\expval{\phi^2}$ together with their power-law fits, versus final temperature for $a=1.0$ (see Table. \ref{tab:Gamma_T}), $g^2/\omega_0^3=0.7$ and $\gamma/\omega_0=0.12$.}
    \label{fig:gamma_a1}
\end{figure}
During the second stage of dynamics, we observe a gradual enhancement of the NFL behavior for fermions as the systems cools down which has already been observed in purely fermionic SYK models in the past \cite{zhang-syk-2019,cheipesh2021}. In addition, phonons are softened over time toward their low temperature gapless state, an exclusive property of the Yukawa-SYK model.

Shortly after the oscillatory behavior of the energy is over, the energy starts to relax exponentially to its final value (inset of Fig. \ref{fig:E(t)})
\begin{equation}
    E(t)\approx E_f+A_E e^{-\Gamma_E t}.
\end{equation}
The energy relaxation rate $\Gamma_E$ depends on the final temperature and has a power-law scaling with $T_f$ (Fig. \ref{fig:gamma_a1}). 

The effective temperature also follows a monotonic decrease during the second stage. We see that the late stage relaxation follows an exponential trend
\begin{equation}
   T_\mathrm{eff}(t)\approx T_f + A_T e^{-\Gamma_Tt}.
\end{equation}
The temperature relaxation rate $\Gamma_T$ (Fig. \ref{fig:gamma_a1}) turns out to be close to the energy relaxation rate $\Gamma_E$ for different values of $a$ (Table. \ref{tab:Gamma_T}) and has the same scaling with $T_f$ as $\Gamma_E$ given by Eq. (\ref{Gamma_E_an}) below.

The exponential relaxation is not exclusive to total energy and temperature. The fluctuations of phonon displacement $\expval{\phi^2}$ also relax exponentially to their final value, with the same rate given above (Fig. \ref{fig:gamma_a1}). Moreover, total energy is proportional to $\expval{\pi^2}$ according to Eq. (\ref{simple_energy}) and therefore, this quantity has the same relaxation profile. The fact that all of these quantities relax in the same way as temperature suggests that the dynamics of the system is completely captured by the effective temperature. This is consistent with the general picture of quantum critical systems, where temperature is the only relevant energy scale \cite{sachdev_2011}. We will confirm this hypothesis in the next section and find the dependence of $\Gamma$ on $T_f$ as
\begin{equation}\label{Gamma_E_an}
    \Gamma_E\sim T_f^{a-4\Delta+1},
\end{equation}
where $a$ is the exponent of the phonon bath defined in Eq. (\ref{ph_bath_spec}) and $\Delta$ is given by Eq. (\ref{Delta}). We see that the exponent is always positive for an irrelevant bath according to Eq. (\ref{irrel_cond}). 

We can evaluate the efficiency of the bath to thermalize the critical system by looking at Eq. (\ref{Gamma_E_an}). As the bath becomes less relevant in the IR limit, the relaxation rate gets suppressed at small temperatures. The limit $a=4\Delta-1$ when the exponent of $T_f$ in Eq. (\ref{Gamma_E_an}) becomes zero corresponds to a marginal system bath coupling.  
\begin{table}[t]
    \centering
    \begin{tabular}{|c|c|c|}
    \hline
    Bath exponent & Numerics & Analytics ($a+1-4\Delta$) \\
    \hline
    $a=1$ & $0.32$ & $0.32$ \\
    \hline
    $a=1.2$ & $0.53$ & $0.52$ \\
    \hline
    $a=0.9$ & $0.21$ & $0.22$ \\
    \hline
    \end{tabular}
    \caption{Comparison of the exponent of temperature relaxation rate $x$ defined as $\Gamma_T\propto T_f^x$ found from the numerical solution of quantum kinetic equations and the analytical hydrodynamical approximation. The parameters $a$ and $\Delta$ are respectively defined in Eqs. (\ref{ph_bath_spec}) and (\ref{Delta}).}
    \label{tab:Gamma_T}
\end{table}

\subsection{Hydrodynamical approximation}\label{sec:hydro}
As shown before, all physical observables relax with the same rate in the slow stage of thermalization. This can happen, for instance, if all of these quantities could be uniquely determined by only one of them such as the temperature. The SYK model and its variants are all-to-all connected interacting models which can efficiently redistribute energy \cite{Maldacena-prd16,eberlein,knapp-prl20,schiro-prb22}. When we couple these systems to thermal baths, we expect the energy transfer between the system and the bath to be the slowest relevant process and thus, determining the rate of relaxation. At every instant of time, the system is in local equilibrium and all of the observables are given by their value in equilibrium at temperature $T_\mathrm{eff}(t)$. This situation is similar to the "hydrodynamic" relaxation of translationally invariant initial states. Despite the similarity, there is an important difference between our setup and those studied by Refs. \cite{Mukerjee_2006,Lux_2014,Bohrdt_2017} in which the energy is locally conserved and hence, is described by a stochastic diffusion equation after local equilibrium has been established. The scale invariance of the diffusion equation results in the power-law decay of observables at long times, whereas in our system, the energy is not conserved due to coupling to the bath and the long time decay is exponential, except for when the bath is at zero temperature (see the end of this section). Possibly, our case is closer to the hydrodynamic regimes discussed in Refs. \cite{Bouchoule_2020,Bastianello_2021} where collisions are faster than losses, although, these systems are integrable in the absence of losses in contrast to the Yukawa-SYK model.

To check the validity of such "hydrodynamic" hypothesis, we assume the system to be in thermal equilibrium at temperature $T_\mathrm{eff}(t)$ and find energy transfer rate and total energy in terms of $T_\mathrm{eff}(t)$. We solve this closed set of equations to find the relaxation profile of the temperature and compare the results to the numerics of Sec. \ref{sec:second_stage}. The energy transfer rate between the system and the bath can be found in terms of the Green's functions of phonons in the system and the bath (see Appendix \ref{appen_edot})
\begin{multline}\label{edot_neu}
    \partial_t E(t) = -\frac{iN}{2} \hspace{-2pt} \int_0^t \hspace{-3pt} \left(\mathfrak{D}^K(t',t) \partial_t D^A(t',t) \right. \\ \left. +\,\mathfrak{D}^R(t,t')\partial_t D^K(t',t) \right)dt'.
\end{multline}
When the system is in a quasi-thermal state with the slowly varying temperature $T_\mathrm{eff}(t)$, we can assume time translation symmetry and use the expressions for functions appearing in Eq. (\ref{edot_neu}) at thermal equilibrium to get
\begin{multline}\label{edot_eu}
    \partial_t E=N \int \frac{d\omega}{2\pi} \omega J(\omega)\, \mathrm{Im}\,D^R(\omega) \\ \times \left\{ \coth{\left(\frac{\omega}{2T_\mathrm{eff}(t)}\right)}-\coth{\left(\frac{\omega}{2T_f}\right)}\right\} ,
\end{multline}
where $J(\omega)$ was given by Eq. (\ref{ph_bath_spec}). At low temperatures, we can employ the scaling form of the phonon Green's function \cite{ysyk2019} to find
\begin{equation}\label{edot_scaling}
    \partial_t E=-N\gamma g^{2(4\Delta-1)}\omega_0^{-8\Delta} T_\mathrm{eff}^{3+a-4\Delta}f(\frac{T_\mathrm{eff}}{T_f}),
\end{equation}
where $f(x)$ is given by
\begin{multline}
    f(x)=-\frac{2^{1+a} \pi \cos{(2\pi \Delta)}}{\cos{(\pi \Delta)}\sin^3{(\pi \Delta)}\Gamma(2\Delta)\Gamma(1-4\Delta)}\\ \times \qty(\frac{2\Delta-1}{8\Delta^2 \sin{\frac{\pi}{2\Delta}}})^{4\Delta} \int_0^{\infty} y^{a+2-4\Delta}\qty(\coth{y}-\coth{yx})\,dy,
\end{multline}
and it has the following limiting behaviors
\begin{equation}
    \lim_{x\to 1}f(x)=0, \qquad \lim_{x\to \infty} f(x)=\mathrm{const}.
\end{equation}
For finite $T_f$ and $T_\mathrm{eff}(t)\gtrsim T_f$, Eq. (\ref{edot_scaling}) gives
\begin{equation}\label{Edot_linear}
    \partial_tE\propto T_f^{2+a-4\Delta}(T_\mathrm{eff}-T_f).
\end{equation}
For a zero temperature bath or when $T_f \ll T_\mathrm{eff} \lesssim g^2/\omega_0^2$, which corresponds to the intermediate stage of the evolution of the system coupled to a finite temperature but sufficiently cold bath, we find
\begin{equation}\label{edot_Tzero}
    \partial_t E\propto T_\mathrm{eff}^{3+a-4\Delta}.
\end{equation}
To identify a closed differential equation for the time-evolution of the temperature, we note that similar to the SYK model \cite{Maldacena-prd16}, the Yukawa-SYK model has a linear specific heat at small temperatures
\begin{equation}\label{E_euc}
    E(T)-E(0)\approx \frac12 N c T^2,
\end{equation}
where $c$ is a non-universal parameter which can be evaluated numerically. The linear specific heat is a consequence of the reparameterization symmetry of the Yukawa-SYK model in the infrared limit \cite{Kim-prb20} which is spontaneously broken  by the saddle point solution in Eqs. (\ref{G_euc}) and (\ref{D_euc}). The degeneracy of the resulting gapless Goldstone modes is lifted by the contributions from UV modes, resulting in a Schwarzian term in the effective action of the fluctuations around the saddle point. Since the irrelevant bath does not affect the low energy physics of the system, we expect a Schwarzian term to be present in the low energy action of a system coupled to bath. The Schwarzian results in a linear specific heat \cite{Maldacena-prd16} at small temperatures. 

 By combining (\ref{edot_scaling}) and (\ref{E_euc}) we get a kinetic equation that governs the time-variation of the temperature:
\begin{equation}\label{Tdot_eu}
    \frac{dT_\mathrm{eff}}{dt}= - \frac{\gamma}{c} g^{2(4\Delta-1)}\omega_0^{-8\Delta}T_\mathrm{eff}^{2+a-4\Delta}f(\frac{T_\mathrm{eff}}{T_f}).
\end{equation}
As a result, for the regime $T_\mathrm{eff}(t)\gtrsim T_f$ the evolution of temperature is given by
\begin{equation}
    \frac{dT_\mathrm{eff}}{dt}\propto - \gamma T_f^{1+a-4\Delta}(T_\mathrm{eff}-T_f).
\end{equation}
Therefore, $T$ relaxes exponentially to $T_f$ with a rate satisfying (\ref{Gamma_E_an}).
The similarity between $\Gamma_E$ and $\Gamma_T$ is now clear from (\ref{Edot_linear}).
\begin{figure}
    \centering
    \includegraphics[width=0.48\textwidth]{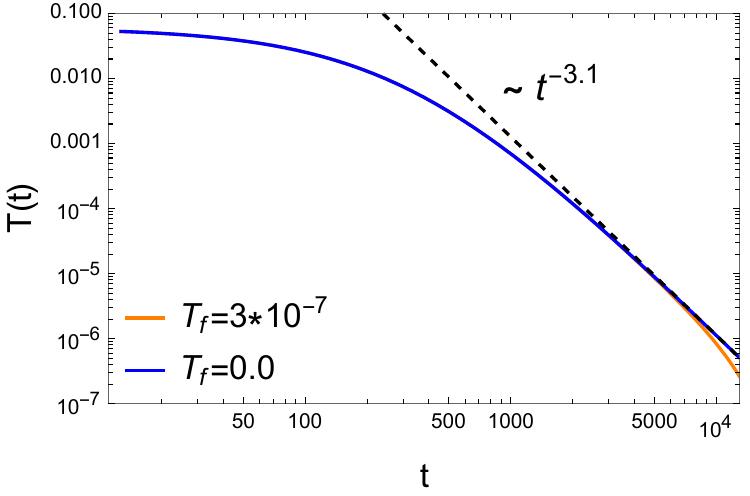}
    \caption{Numerical solution of Eq. (\ref{Tdot_eu}) for the two cases of zero and finite temperature Ohmic baths ($a=1$). The late time power-law decay of temperature is evident for $T_f=0$. For a finite temperature bath, the power-law regime can exists only transiently and it eventually turns into an exponential relaxation at longer times.}
    \label{fig:Tdot_integ}
\end{figure}

In the limit of a zero or extremely low temperature bath in Eq. (\ref{edot_Tzero}), we get a power-law decay for temperature
\begin{equation}\label{powerlaw_T}
    \frac{dT_\mathrm{eff}}{dt}\propto -  T_\mathrm{eff}^{2+a-4\Delta} \to \lim_{t\to \infty}T_\mathrm{eff}(t) \propto t^{-\frac{1}{1+a-4\Delta}}.
\end{equation}
The result of the numerical evaluation of Eq. (\ref{Tdot_eu}) is given in Fig. \ref{fig:Tdot_integ} for baths at zero and finite but small temperatures. As it can be seen, the regime of power-law relaxation of temperature can be difficult to access for a finite temperature bath, as the system may enter the exponential relaxation regime before the power-law behavior can emerge. This is the case for the numerical data presented in this paper as the limited numerical resources prevented us to resolve temperatures which are small enough to observe the power-law decay in Eq. (\ref{powerlaw_T}).

\section{Conclusions}\label{sec:conclusions}
In this work, we studied the quench dynamics of a variant of the Sachdev-Ye-Kitaev (SYK) model with electron-phonon interactions coupled to an external bath. Based on scaling analysis and numerical evaluation of quantum kinetic equations, we showed that for couplings to a generic phonon bath described by the Caldeira-Leggett model, the critical behavior of the system is unaffected. Furthermore, we observed that the system relaxes quickly at short-time/high-frequency scales while global thermalization, corresponding to the equilibration of small-frequency modes, takes longer. The system exhibits quasi-equilibrium behavior at a time-dependent effective temperature obtained from the fluctuation-dissipation theorem. Using the quasi-equilibrium state of the system, we provided an analytical description of the relaxation profile of total energy and temperature in agreement with our numerics. We found that while phonons are directly coupled to the bath, fermions have a lower temperature due to strong correlations between the two species in the NFL phase, while the opposite is true in the FL state.  

There are multiple directions to pursue in the context of the dynamics of what we generally call fermion-boson SYK (FB-SYK) systems. One clear extension of our work is to study quenches in the presence pairing interactions, i.e., when the real and imaginary parts of $g_{ij,k}$ in Eq. (\ref{H-ysyk}) have different second moments. This is the direction that we are currently following. One could also study the evolution of FB-SYK systems under the influence of an external periodic drive. The driving field can be coupled to fermions (similar to Ref. \cite{knapp-prl20} for the SYK model) or phonons. Furthermore, phonons can be driven linearly or parametrically with the possibility of different qualitative and quantitative behaviors. In the superconducting phase, one may investigate the destruction or possibly the transient amplification \cite{babadi_prb16,babadi_prb17} of superconducting correlations in a system with pairing of incoherent fermions.

\emph{Note added} -- During the submission process of this paper, we became aware of a recent work in Ref. \cite{grunwald2023} where the quench dynamics of an isolated superconducting Yukawa-SYK model is studied using Keldysh field theory.

\begin{acknowledgments}
This work was supported by the Deutsche Forschungsgemeinschaft (DFG, German Research Foundation) through TRR 288 - 422213477 (projects B09 and A07) and by the Dynamics and Topology Centre funded by the State of Rhineland Palatinate and Topology Centre funded by the State of Rhineland Palatinate. The authors gratefully acknowledge the computing time granted on the supercomputer MOGON 2 at Johannes Gutenberg-University Mainz (hpc.uni-mainz.de).
\end{acknowledgments}

\nocite{*}

\appendix

\begin{widetext}
\section{Quantum Kinetic Equations}\label{appen_kinetic}
We start from an alternative expression for the free phonon action using Legendre transformation
\begin{equation}\label{s_phpi}
    S_{\mathrm{ph}}=\sum_k^N \oint_{\mathcal{C}}dt_c \, \left(\pi_k \partial_{t_c}\phi_k - H(\pi_k,\phi_k) \right) =\frac12 \sum_k^N \oint_{\mathcal{C}}dt_c \, \Phi_k^T \cdot \mathbf{D_0}^{-1}\cdot \Phi_k,
\end{equation}
where $\Phi_k=(\phi_k, \pi_k)^T$ and 
\begin{equation}
    \mathbf{D_0}^{-1}=\begin{pmatrix}
        -\omega_0^2 & -\partial_{t_c} \\ \partial_{t_c} & -1
    \end{pmatrix},
\end{equation}
The Green's function matrix is defined as
\begin{equation}\label{phi_m}
    \mathbf{D}(t_c,t_c')\equiv \begin{pmatrix}
        D(t_c,t'_c) &  \tilde{B}(t_c,t'_c) \\  B(t_c,t'_c) &  C(t_c,t'_c)
    \end{pmatrix},
\end{equation}
where $D$, $B$ and $C$ were defined in Eqs. (\ref{D(t)}), (\ref{B(t)}) and (\ref{C(t)}). We also have $\tilde{B}(t_c,t_c')\equiv B(t'_c,t_c)$ where we have avoided using the transpose sign $B^T$ as superscripts are reserved for Keldysh indices. Then, the matrix form of Dyson equation in (\ref{dyson_ph}) is given by
\begin{equation}\label{dyson_ph_m}
    \mathbf{D_0}^{-1}. \mathbf{D}(t_c,t_c')=\mathbb{1} \delta (t_c,t_c') + \oint \mathbf{\Pi}(t_c,\tau_c) cdot\mathbf{D}(\tau_c,t_c')\, d\tau_c.
\end{equation}
The self-energy matrix has only one non-zero entry
\begin{equation}\label{ph_se_m}
    \mathbf{\Pi}(t_c,t_c')=\Pi(t_c,t'_c)\begin{pmatrix}
        1 & 0 \\ 0 & 0
    \end{pmatrix},
\end{equation}
Where $\Pi(t_c,t'_c)$ is given in (\ref{total_ph_self}). Putting (\ref{phi_m}) and (\ref{ph_se_m}) in (\ref{dyson_ph_m}) gives
\begin{align}
    \partial_{t_c}D(t_c,t'_c)&=B(t_c,t'_c), \label{dtD}\\
    \partial_{t_c}B(t_c,t'_c)&=-\delta(t_c,t_c')-\omega_0^2 D(t_c,t_c') - \oint \Pi(t_c,\tau_c) D(\tau_c,t'_c)\, d\tau_c,\\
    \partial_{t_c}B^T(t_c,t'_c)&=C(t_c,t'_c),\\
    \partial_{t_c}C(t_c,t'_c)&=-\delta(t_c,t'_c) -\omega_0^2 \tilde{B}(t_c,t'_c) - \oint \Pi(t_c,\tau_c) \tilde{B}(\tau_c,t'_c)\, d\tau_c. \label{dtC}
\end{align}
We use Langreth rules \cite{Langreth1976} and write Eqs. (\ref{dyson_f}) and (\ref{dtD})-(\ref{dtC}) together with their Hermitian conjugates in terms of greater, lesser, retarded and advanced functions to get quantum kinetic equations (QKE)
\begin{align}
    +i\partial_t G^\gtrless(t,t')&= \int \left( \Sigma^R(t,\tau) G^\gtrless (\tau,t') + \Sigma^\gtrless(t,\tau)G^A(\tau,t'') \right)\, d\tau,\\
    -i\partial_{t'} G^\gtrless(t,t')&= \int \left( G^R(t,\tau) \Sigma^\gtrless (\tau,t') + G^\gtrless(t,\tau)\Sigma^A(\tau,t'') \right)\, d\tau,\\
    \Sigma^\gtrless(t,t')&=ig^2 G^\gtrless(t,t')D^\gtrless(t,t'), \\
    \partial_t D^\gtrless (t,t')&=B^\gtrless (t,t'), \\
    \partial_{t'} D^\gtrless (t,t')&=\tilde{B}^\gtrless(t,t'), \\
    \partial_{t'}B^\gtrless(t,t')&= \partial_t \tilde{B}^\gtrless(t,t') = F^\gtrless(t,t'), \\
    \partial_t B^\gtrless(t,t')&=-\omega_0^2 D^\gtrless(t,t') - \int \left( \Pi^R(t,\tau) D^\gtrless (\tau,t') + \Pi^\gtrless(t,\tau)D^A(\tau,t'') \right)\, d\tau,\\
    \partial_t F^\gtrless (t,t')&=-\omega_0^2 \tilde{B}^\lessgtr(t',t)-\int \left( \Pi^R(t,\tau) \tilde{B}^\lessgtr (t',\tau) + \Pi^\gtrless(t,\tau)\tilde{B}^A(\tau,t'') \right)\, d\tau,\\
    \partial_{t'} F^\gtrless (t,t')&=-\omega_0^2 B^\lessgtr(t',t)-\int \left( B^R(t,\tau) \Pi^\lessgtr (t',\tau) + B^\gtrless(t,\tau)\Pi^A(\tau,t'') \right)\, d\tau, \\
    \Pi^\gtrless(t,t')&=-2ig^2 G^\gtrless(t,t')G^\lessgtr(t',t) + \mathfrak{D}^\gtrless(t,t').
\end{align}
The phonon self-energy contains the contribution from the bath given by (\ref{total_ph_self}). The retarded and advanced functions are defined according to Eqs. (\ref{ret_def}) and (\ref{adv_def}). Note that $\tilde{B}^A(t,t')\neq \left[B^A(t,t')\right]^T$ but 
\begin{equation}
    \Tilde{B}^A(t,t')=\Theta(t'-t)\left( \Tilde{B}^<(t,t')-\Tilde{B}^>(t,t') \right)=\Theta(t'-t)\left( B^>(t',t)-B^<(t',t) \right)=B^R(t',t),
\end{equation}
therefore, the causality of kinetic equations is respected.

In order to numerically solve QKE, we pay attention to the causal structure in (\ref{causal_cond}) and use an implicit mid-point method on an $N\times N$ grid with step size $dt$ (see Fig. \ref{causal_fig}). To access temperature $T$, we should be able to resolve frequencies comparable to $T$. Consequently, the grid size should satisfy the condition $N\,dt \gtrsim \pi T^{-1}$. Furthermore, the step size should be small enough to prevent instabilities of the solution at late stages of the evolution. The main numerical cost comes from increasing the grid size $N$. We tested two choices of $N=3000,\, dt=0.1$ and $N=5000,\, dt=0.05$ and the results were in agreement.
\begin{figure}[t]
    \centering
    \includegraphics[width=0.3\textwidth]{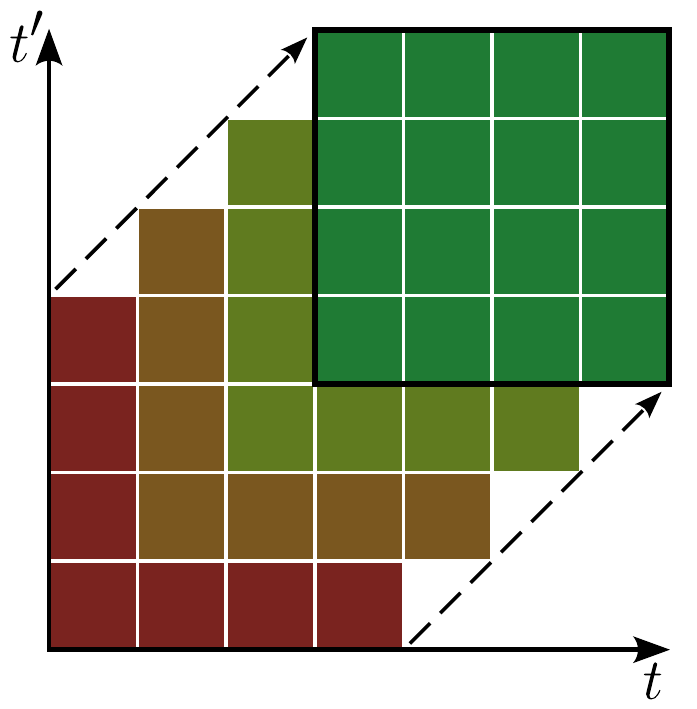}
    \caption{Illustration of how QKE are solved in the 2 dimensional time space.}
    \label{causal_fig}
\end{figure}
\end{widetext}

\section{Total energy}\label{app:total_energy}
The total energy is given by the expectation value of (\ref{H-ysyk}). The contribution of the free phonon term is given by:
\begin{equation}\label{E_ph}
    \langle H_{\mathrm{Ph}}(t) \rangle=\frac{i}{4}N(-\partial_t^2+\omega_0^2)D^K(t,t')|_{t'\to t}.
\end{equation}
To find the contribution of the interaction term we add the source term $-i\oint J(t) H_{int}(t)\, dt$ to the action and take the functional derivative with respect to the source
\begin{equation}
    \overline{\langle H_{int}(t) \rangle} =i\frac{\delta}{\delta J(t)}\left( \int \mathcal{D}[\psi,\bar{\psi},\phi]\, \overline{e^{iS-i\oint J(t) H_{int}(t) \,dt}} \right).
\end{equation}
Evaluating the functional derivative results in
\begin{equation}\label{E_int_DGG}
    \overline{\langle H_{int}(t) \rangle}=Ng^2 \oint D(t,t'_c)G(t,t'_c)G(t'_c,t)\, dt'_c.
\end{equation}
To simplify the expression for total energy, we note that (\ref{E_int_DGG}) can be written in terms of phonon self-energy given in (\ref{ph_self}) to get
\begin{equation}
    \overline{\langle H_{int}(t) \rangle}=\frac{i}{2}N\oint \Pi_{\mathrm{YSYK}}(t,t'_c)D(t'_c,t)\, dt'_c,
\end{equation}
Where $\Pi_{\mathrm{YSYK}}$ is the contribution of Yukawa-SYK interaction to phonon self-energy, excluding the coupling to the bath given by $\mathcal{D}(t,t')$. In the next step, we make use of SD equation for phonons (\ref{dyson_ph}) to write
\begin{multline}\label{E_int}
    \overline{\langle H_{int}(t) \rangle}=-\frac{i}{4}N(\partial_t^2+\omega_0^2)D^K(t,t')|_{t'\to t}\\-\frac{i}{2}N \oint \mathfrak{D}(t,t'_c)D(t'_c,t)\, dt'_c.
\end{multline}
The total energy is found by adding (\ref{E_ph}) to (\ref{E_int}), resulting in the cancellation of the term proportional to $\omega_0^2$
\begin{multline}
    \frac{E(t)}{N}=-\frac{i}{2}\partial_t^2D^K(t,t')|_{t'\to t}\\-\frac{i}{4}\int \left[\mathfrak{D}^R(t,t')D^K(t',t)+\mathfrak{D}^K(t,t')D^A(t',t)\right]\, dt'.
\end{multline}
The first term can be written as $\langle \pi^2(t)\rangle$ which was mentioned in Eq. (\ref{simple_energy}) of the main text.

\section{Energy transfer rate}\label{appen_edot}
The energy current operator is given by
\begin{equation}
    \partial_t H_S=i[H_S+H_B+H_{SB},H_S]=i[H_{SB},H_S],
\end{equation}
where $H_S$ is the Hamiltonian of an isolated system given by (\ref{H-ysyk}) and $H_{SB}$ is defined in (\ref{hsb_phon}). After calculating the commutator we get
\begin{equation}
    \partial_t H_S=-\sum_l X_l \dot{\phi}_l.
\end{equation}
The expectation value of $X_l\dot{\phi}_l$ reads as
\begin{equation}
   \langle X_l(t)\dot{\phi}_l(t)\rangle=\oint \mathfrak{D}(t,t')\langle \phi_l(t')\dot{\phi}_l(t)\rangle\, dt',
\end{equation}
where $\mathfrak{D}(t,t')$ is the contour-ordered Green's function of the bath defined in (\ref{s_b_ph}). Therefore, the energy transfer rate is given by
\begin{multline}
    \partial_t E(t)\equiv\langle \partial_t H_S \rangle = -iN \oint \mathfrak{D}(t,t')\partial_t D(t',t)\,dt' \\ = -\frac{iN}{2} \hspace{-2pt} \int_0^t \hspace{-3pt} \left(\mathfrak{D}^K(t',t) \partial_t D^A(t',t)+\mathfrak{D}^R(t,t')\partial_t D^K(t',t) \right)dt',
\end{multline}
which is the quoted result in Eq. (\ref{edot_neu}) of the main text.

\section{The Fermi Liquid Electron-Phonon System}\label{app:FL}
\begin{figure}[!t]
    \centering
    \includegraphics[width=.48\textwidth]{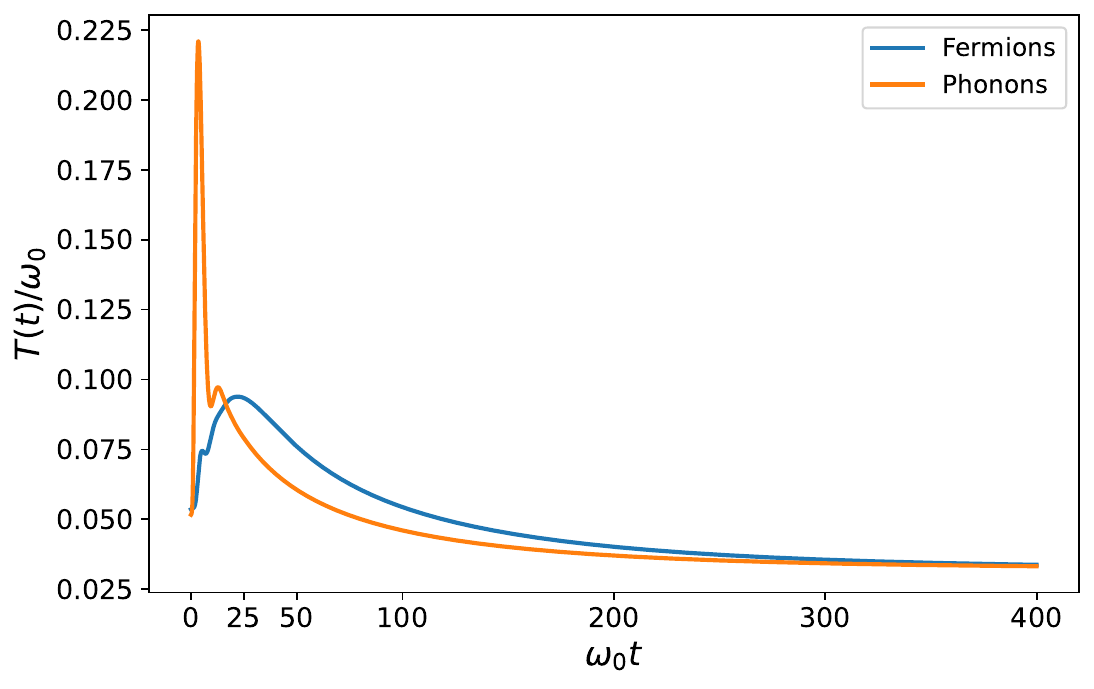}
    \caption{Effective temperatures for a Fermi liquid Yukawa-SYK system coupled to an external Ohmic bath. The parameters are $t/\omega_0=1.0$, $g^2/\omega_0^3=0.7$ and $\gamma/\omega_0=0.2$.}
    \label{fig:T(t)_randhop}
\end{figure}
\begin{figure}[!t]
    \centering
    \includegraphics[width=.48\textwidth]{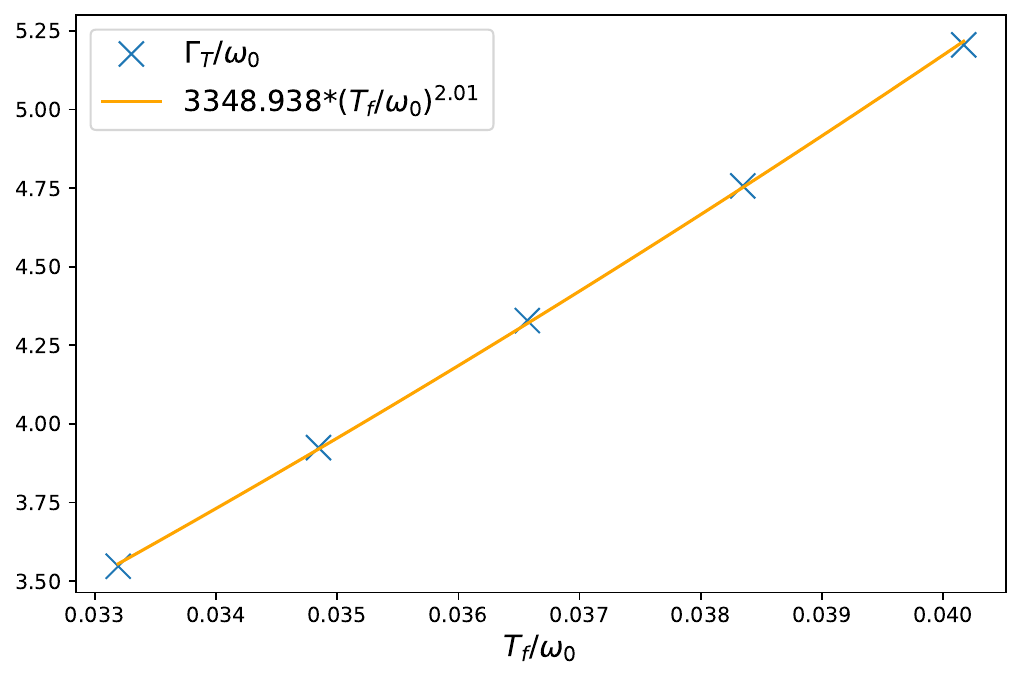}
    \caption{Temperature relaxation rate for a Fermi liquid Yukawa-SYK system coupled to an external Ohmic bath. The parameters are $t/\omega_0=1.0$, $g^2/\omega_0^3=0.7$ and $\gamma/\omega_0=0.2$.}
    \label{fig:rand_hop_GammaT}
\end{figure}
Fermi liquid behavior can be obtained by adding a random hopping term $H_t$ to the Yukawa-SYK Hamiltonian in Eq. (\ref{H-ysyk})
\begin{equation}
    H_t=-\frac{1}{\sqrt{N}}\sum_{ij,\sigma}t_{ij} \psi^\dagger_{i\sigma}\psi_{j\sigma}.
\end{equation}
The hopping amplitude $t_{ij}=t'_{ij}+it''_{ij}$ is a random Gaussian variable and satisfies
\begin{align}
    \overline{t_{ij}}&=0, \\
    \overline{t'_{ij}t'_{nm}}&=\frac{t^2}{2} \qty(\delta_{in}\delta_{jm}+\delta_{im}\delta_{jn}), \\
    \overline{t''_{ij}t''_{nm}}&=\frac{t^2}{2} \qty(\delta_{in}\delta_{jm}-\delta_{im}\delta_{jn}).
\end{align}
The hopping term results in an extra contribution to the fermion self-energy in Eq. (\ref{f_self}) given by
\begin{equation}
    \Sigma_t(t_c,t'_c)=t^2 G(t_c,t'_c),
\end{equation}
where in the absence of the Yukawa interaction and at equilibrium yields the following fermion Green's function
\begin{equation}\label{GR_rand_hop}
    G^R(\omega)=\frac{2}{\omega+i\sqrt{4t^2-\omega^2}}.
\end{equation}
According to the scaling dimension of fermion operators $[\psi]=\Delta$ given by Eq. \ref{G_euc}, $H_t$ is a relevant operator near the Yukawa-SYK fixed point and we can treat the Yukawa interaction as a perturbation to randomly hopping fermions and free phonons. The Yukawa contribution to the imaginary parts of the fermion and phonon self-energies at equilibrium reads
\begin{multline}\label{rand_hop_f_se}
    \mathrm{Im}\, \Sigma_{g}^R(\omega)=-g^2 \int \mathrm{Im}\,G^R(\nu)\, \mathrm{Im}\,D^R(\omega-\nu) \\ \times \qty[\tanh{\frac{\nu}{2T}}-\coth{\frac{\nu-\omega}{2T}}]\, \frac{d\nu}{2\pi},
\end{multline}
\begin{multline}\label{rand_hop_p_se}
    \mathrm{Im}\, \Pi_g^R(\omega)=-2g^2 \int \mathrm{Im}\,G^R(\nu)\,\mathrm{Im}\,G^R(\nu+\omega)  \\ \times \qty[\tanh{\frac{\nu+\omega}{2T}}-\tanh{\frac{\nu}{2T}}]\, \frac{d\nu}{2\pi}.
\end{multline}
We substitute $G^R$ in Eq. (\ref{rand_hop_p_se}) from Eq. (\ref{GR_rand_hop}). After putting the contribution from the coupling to the bath and the Yukawa interaction together we find
\begin{equation}\label{rand_hop_p_se_fin}
    \mathrm{Im}\, \Pi^R(\omega)\approx -\gamma \sin{\qty(\frac{\pi a}{2})} \,\mathrm{sgn}\,(\omega) |\omega|^{a} - \frac{2g^2}{\pi t^2} \omega.
\end{equation}
As expected for a fermionic system with a smooth DOS at the Fermi energy, the interaction of phonons with fermion charge fluctuations results in the Landau damping of phonons. Depending on the exponent of the bath $a$ and temperature, phonons exhibit different relaxation behaviors. For Ohmic and sub-Ohmic baths ($a\le 1$) the Yukawa interaction does not alter the dynamics of phonons at temperatures below the crossover scale $\omega^\star \sim (\gamma t^2 /g^2)^{1/(1-a)}$ while for a super-Ohmic bath the Yukawa self-energy dominates the spectrum of phonons below $\omega^\star$. Typically, the system-bath coupling $\gamma$ is a small parameter and the Yukawa vertex determines the relaxation of phonons down to very small temperatures. Note that in contrast to the SYK regime ($t\to 0$), phonons are not critical at $T\to 0$ and have a finite renormalized gap $\omega_r$; therefore, for small energies we have $\mathrm{Im}\, D^R(\omega)\approx \frac{1}{\omega_r^4} \mathrm{Im}\,\Pi^R(\omega)$. Substituting Eq. (\ref{rand_hop_p_se_fin}) into Eq. (\ref{rand_hop_f_se}) yields
\begin{equation}
    \mathrm{Im}\, \Sigma_g^R(\omega)\approx -\frac{g^2}{2\pi \omega_r^4 t}\qty(\gamma \sin{\qty(\frac{\pi a}{2})} \abs{\omega}^{a-1}+\frac{2g^2}{\pi t^2})\abs{\omega}^2.
\end{equation}
Albeit the scaling of the fermion self-energy depends on the spectrum of the bath, we have $\Sigma(\omega)/\omega\xrightarrow{\omega \to 0}0$ and therefore, the scattering rate of fermions is consistent with the Fermi liquid picture. 

By solving the kinetic equations for this system numerically, we can employ Eq. (\ref{T_eff}) to find the effective temperatures for fermions and phonons. As it can be seen in Fig. \ref{fig:T(t)_randhop}, phonons are colder than fermions during intermediate and late stages of the evolution as they are directly coupled to the bath, in contrast to the NFL phase discussed in the main text. Furthermore, we observe a late-time exponential relaxation of temperature $T-T_f \sim e^{-\Gamma_T t}$. However, the scaling of the relaxation rate $\Gamma_T$ is different from the one for the critical system in Eq. (\ref{tab:Gamma_T}). For an Ohmic bath ($a=1$), numerics show $\Gamma_T\propto T^2$ (see Fig. \ref{fig:rand_hop_GammaT}) while for a super-Ohmic bath with $a=1.2$ we get $\Gamma_T\propto T^{2.25}$. We can use the hydrodynamical approximation for the energy transfer rate in Eq. (\ref{edot_eu}) together with a linear specific heat for the Fermi liquid to get
\begin{equation}
    \Gamma_T \propto T^{1+a},
\end{equation}
which is consistent with the results given above. The important observation here is that the relaxation of the FL is much slower than the SYK phase at small temperatures.
\bibliography{refs}

\end{document}